\documentclass[twocolumn,superscriptaddress,pre,nobalancelastpage]{revtex4}

\usepackage{amsmath}
\usepackage{amsbsy}
\usepackage{mathtools}
\usepackage{bbm}
\usepackage{dcolumn}
\usepackage{enumitem}
\usepackage{graphicx}
\usepackage{float}
\usepackage{xcolor}
\usepackage{comment}
\usepackage{hyperref}

\newcommand\bm{\boldsymbol}

\usepackage{color}

\newcommand{\change}[1]{{\color{black}{#1}}}

\begin{document}

\title{Heterogeneous message passing for heterogeneous networks}

\author{George T. Cantwell}
\affiliation{Santa Fe Institute, 1399 Hyde Park Road, Santa Fe, New Mexico 87501, USA}
\author{Alec Kirkley}
\affiliation{Institute of Data Science, University of Hong Kong, Hong Kong}
\affiliation{Department of Urban Planning and Design, University of Hong Kong, Hong Kong}
\affiliation{Urban Systems Institute, University of Hong Kong, Hong Kong}
\author{Filippo Radicchi}
\affiliation{Center for Complex Networks and Systems Research, School of Informatics, Computing, and Engineering, Indiana University, Bloomington, Indiana 47408, USA}

\begin{abstract}
  Message passing (MP) is a computational technique
  used to find approximate solutions to a variety of problems defined on networks.
 MP approximations are generally accurate in locally tree-like
 networks but require corrections to maintain their accuracy level in
 networks rich with short cycles.
However, MP may already be computationally challenging on very large networks
and additional costs incurred by correcting for cycles could be prohibitive.
We show how the issue can be addressed.
By allowing each node in the network to have its own level of
approximation, one can focus on improving the accuracy of MP
approaches in a targeted manner.
We perform a systematic analysis of $109$ real-world networks
and show that our node-based MP approximation is
able to increase both the accuracy and speed of traditional MP
approaches. We find that,
compared to conventional \change{MP},
a heterogeneous approach based on a simple heuristic
is more accurate in 81\% of tested networks, faster in 64\% of
cases, and both more accurate and faster in 49\% of cases.
\end{abstract}

\maketitle

\section{Introduction}

Message passing (MP), sometimes called belief propagation or the cavity
method, is a computational technique aimed at
solving problems or characterizing
processes on networks~\cite{newman2022message}.
Examples
include
spreading processes~\cite{karrer2010message,
  lokhov2014inferring, lokhov2015dynamic, altarelli2013optimizing, altarelli2014containing, bianconi2021message}, community
detection~\cite{decelle2011inference, radicchi2018decoding}, sampling
strategies~\cite{radicchi2018uncertainty}, spectral
properties~\cite{dorogovtsev2003spectra, rogers2008cavity}, and
percolation~\cite{karrer2014percolation, hamilton2014tight, radicchi2015breaking, bianconi2017fluctuations, bianconi2018rare}.

MP techniques are closely related to mean-field approximations~\cite{mezard2009information, mackay2003information}, in which one relates a quantity of interest at each node to those of their neighbors.
For example, the event in which someone catches an infectious disease is related to whether the people they are in contact with catch the disease.
A mean-field analysis proceeds by replacing unknown quantities defined
on the nodes of the network with average or expected values, 
%and
(incorrectly!) assuming that these quantities are independent.
One derives a set of self-consistent equations to be solved.
In such mean-field approximations there is one equation for each node in the network.
Unfortunately, these approximations can be inaccurate.

MP approaches follow a similar logic but at least partially account
for correlations induced by edges.
In place of the independence assumption of a mean-field approximation, one makes a \emph{conditional} independence assumption.
As for mean-field approximations, MP approximations introduce
a system of self-consistent equations to be solved.
Now, however,
there are two equations for each \emph{edge} in the network.

Conventional MP techniques are often exact on trees (networks without
cycles), and are justified on general networks by a
locally tree-like assumption.
When
applied to real networks,
MP methods often generate fairly accurate predictions~\cite{melnik2011unreasonable}.
Mistakes in the predictions can be attributed to the inability of the locally tree-like assumptions to account for the correlations introduced by cycles.

However, because many networks have a relatively high density of short cycles it is important to be able to account for them.
Social networks, for instance, typically have large numbers of triangles~\cite{watts1998collective, newman2010networks}.
One mechanism that would give rise to a large number of triangles is
triadic closure---the process whereby two of your friends become
friends with each other. %~\cite{granovetter1973strength}.
Likewise shared familial, vocational, or geographical ties can lead to densely connected subgroups of individuals, and hence large numbers of triangles.

A few attempts
to account for correlations due to short loops
exist in the literature.
Some previous methods, such as those of Refs.~\cite{newman2009random, radicchi2016beyond}, do not generalize to arbitrary combinations of short loops, or suffer from the limitation of being problem specific.
One promising direction is the approach of Cantwell and Newman~\cite{cantwell2019message}, which accounts for correlations caused by arbitrary short loops.

The framework of Ref.~\cite{cantwell2019message} relies on a procedure for constructing appropriately defined neighborhoods around each node.
We will refer to this procedure as the neighborhood message passing
(NMP) approach.
The size of the neighborhoods can be increased in order to improve the accuracy of the predictions, but this increase in accuracy comes at the cost of an increasingly complex set of equations to solve.
The utility of NMP
follows from the fact that the method provides
good results for relatively small neighborhoods.
The accuracy of NMP 
has been demonstrated for bond percolation, spectral properties of sparse matrices, and the Ising model~\cite{cantwell2019message, kirkley2021belief}.

However, many networks have heterogeneous degree distributions
\cite{barabasi1999emergence}, and this property may cause unique
problems.
First, heterogeneous degree distributions can imply a large density of short cycles~\cite{bianconi2005loops}.
In a random graph with \change{$n$ nodes and} degree distribution $p_k$, the expected
number of triangles per node \change{$\langle t \rangle$} is to leading order
\begin{equation}
	\langle t \rangle \propto \frac{\langle k^2 \rangle^3}{2 n \langle k \rangle^3}
\end{equation}
where $\langle k \rangle = \sum_k k p_k$ and $\langle k^2 \rangle = \sum_k k^2 p_k$.
If $\langle k^2 \rangle$ diverges as $n^{1/3}$ or faster, the expected number of triangles per node diverges, even if the network is sparse.

Second, heterogeneous degree distributions may cause MP schemes to be
somewhat more computationally demanding than they are in networks with
homogeneous degree distributions.
By definition, each node of degree~$k$ has $k$ edges.
Each of these edges has a corresponding equation that typically depends on $k-1$ other variables.
Evaluating these equations \change{for a network with $n$ nodes} may thus require $\mathcal{O}(n \langle k^2 \rangle )$ operations. 
When $\langle k^2 \rangle$ is large (or diverging)
the numerical solution of the equations
could be expensive.

Heterogeneous degree distributions may thus cause both accuracy and speed degradation for traditional MP approaches.
As discussed, the NMP approach trades off speed for accuracy in networks with short cycles. 
For networks with relatively homogeneous degree distributions, such as social or biological networks \cite{newman2010networks}, the cost may be quite acceptable.
However, NMP may considerably exacerbate the speed issues caused by heterogeneous degree distributions, and the additional cost may simply be prohibitive.
In this paper, we present a solution to this problem, allowing for accurate and fast approximations for real-world networks with heterogeneous degree distributions.

Our solution embraces heterogeneity and relies on an appropriately heterogeneous approximation.
Large-degree nodes can be approximated by conventional mean-field approximations, since the aggregate fluctuations of their neighbors should be small by the law of large numbers.
Conversely, low-degree nodes can be approximated either by conventional MP, or by NMP when there is a large density of short loops.
By tailoring the level of approximation for each node we can deploy our methods to arbitrary networks.

The remaining sections of the paper are structured as follows.
In Sec.~\ref{sec:neigborhood_het}, we systematically explore the properties of 
%network 
neighborhoods in real-world networks, observing considerable heterogeneity.
\change{In Secs.~\ref{sec:het_NMP} and~\ref{sec:results}, 
we first derive and then test} a
heterogeneous
NMP approach for computing the  spectral properties of real networks.
We find that our approach is able to increase on both the accuracy and
the speed of MP in about $50\%$ of cases.
\change{Finally, in Sec.~\ref{sec:discussion}, we show that the NMP approach can be also used in estimating properties of the zero-field Ising model on networks, and then conclude the paper.}

\section{Neighborhood heterogeneity}
\label{sec:neigborhood_het}

\begin{figure}
  \includegraphics[width=0.3\textwidth]{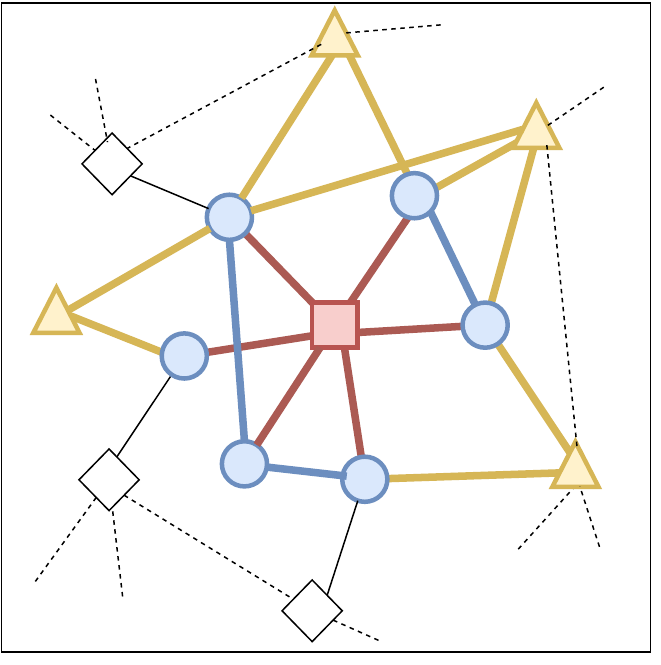}
  \caption{
	{\bf Definition of the $r$-neighborhood of a node.}
        We denote the order $r$ neighborhood of node~$i$ as $E_i^r$.
        Neighborhoods are naturally defined in terms of edges;
        the set of nodes that are in the
        neighborhood, i.e., $N_i^r$, is composed of all nodes at the
        end of at least one edge in $E_i^r$.
          For the focal node $i$, shown as
	  a red square in the figure, the neighborhood $E_i^0$ consists of all edges incident
	  to the node, depicted in red (edges between the square and the
	  circles).  For $E_i^1$, the neighborhood consists of all edges in
	  $E_i^0$ along with all edges between neighbors of $i$, depicted in
	  blue (edges between circles).  For $E_i^2$, the neighborhood consists
	  of all edges in $E_i^1$ along with all edges on paths of length two
	  neighbors of $i$, depicted in yellow (edges between circles and
	  triangles). Nodes in the neighborhoods $N_i^0$ and $N_i^1$
          are denoted by blue circles; $N_i^2$  is composed of all
          nodes denoted as either
          blue circles or yellow triangles.
  }
  \label{fig:diagram}
\end{figure}

In the NMP approach, one sets a value of $r$ for the network.
Given $r \geq 0$, one constructs the neighborhoods of each node,
defined by a set of edges.
Specifically,  the $r$-neighborhood around node~$i$, denoted ${E}_i^r$, consists of all edges incident to node~$i$, along with all paths of length $r$ or shorter between nodes adjacent to~$i$.
An example of this construction is shown in Fig.~\ref{fig:diagram}.
All nodes that appear at the end of edges in ${E}_i^r$
compose the set ${N}_i^r$. 
\change{Note, $r$-neighborhoods are defined by cycles; the neighborhood ${N}_i^r$ is \emph{not} equivalent to the set of nodes that are at distance $r$ from node~$i$.}

To begin, we compute the size of the neighborhoods of $109$ real-world networks. 
Percolation properties of these networks have been
investigated in Refs.~\cite{radicchi2015predicting,
  radicchi2015breaking, radicchi2016beyond}.
The corpus contains networks of different nature, including technological,
biological, social, and information networks.
See Tables~\ref{tab:1}-~\ref{tab:3}
for details.
All networks in the corpus are relatively sparse.
Other structural properties---e.g., size, clustering coefficient, average length, degree distribution and degree correlations---vary greatly within the corpus.

\begin{figure}[!htb]
  \includegraphics[width=0.5\textwidth]{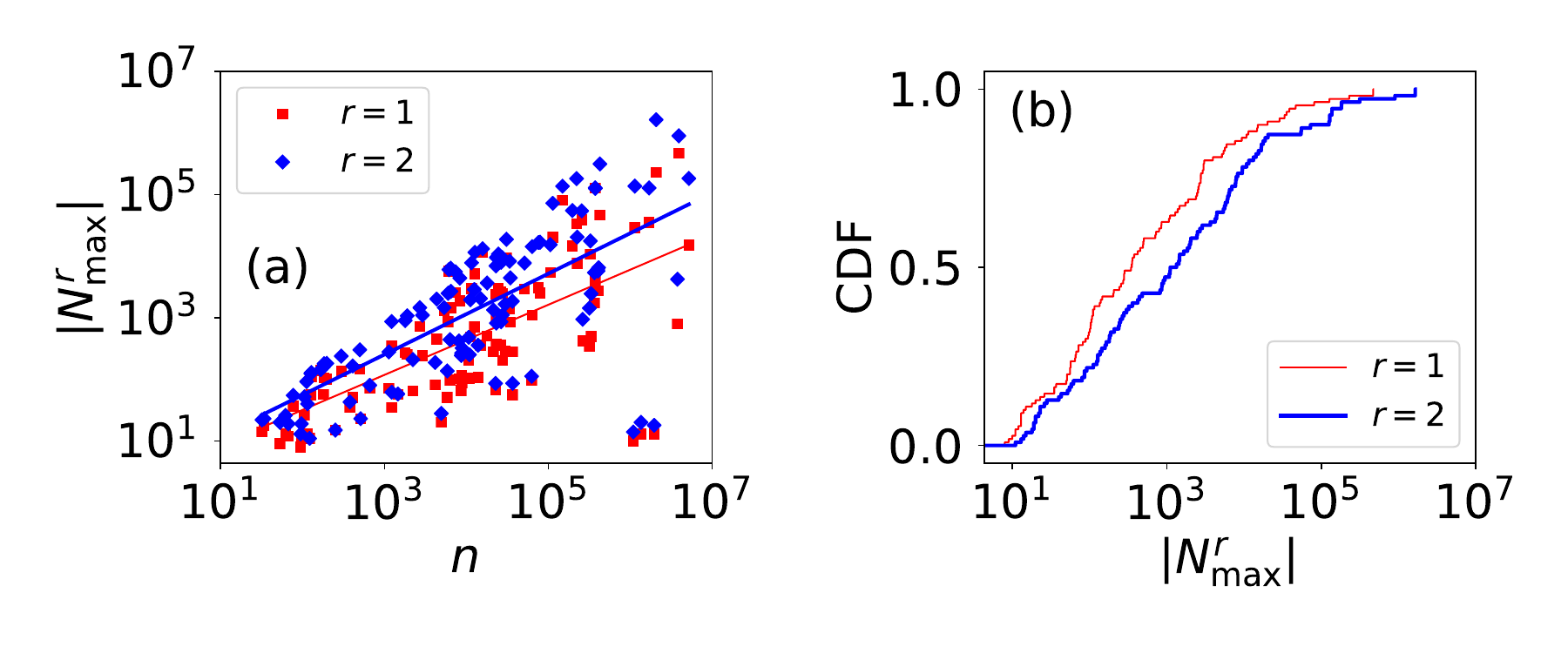}
  \caption{
    {\bf Size of the largest neighborhoods in real-world networks.}
    {\bf (a)} We consider $109$ real-world networks. For
    each network, we evaluate the size of the largest neighborhood $|N_{\max}^r|$
    and plot it as a function of the
    size $n$ of the network. Each point in the plot refers to a network. The
    lines are power-law fits of the data points, i.e., $|N_{\max}^r|
    \sim n^{\beta^r}$. The best fit
    is obtained using linear regression on the logarithms of the
    data. We find $\beta^{1} =0.57$ (Pearson' s linear correlation
    coefficient $0.69$) and $\beta^{2} =0.67$ $(0.71)$.
  {\bf (b)} Cumulative distribution function (CDF) of $\left|
	N^{r}_{\max} \right|$. Data are the same as in panel (a).}
    \label{fig:2}
  \end{figure}

To get a sense of how the NMP approximation will scale, we look at the dependence of the size of the largest neighborhood,
i.e., $|N_{\max}^r| = \max_i | N_{i}^r|$, with the network size.
Our results, reported in Fig.~\ref{fig:2}, indicate that the size of
the largest neighborhood grows with the network size.
There are a few networks for which the largest neighborhood are very small compared to the network size.
This is the case of networks with homogeneous degree distributions and/or 
strong spatial embedding, as for example road networks.
Networks with heterogeneous degree distributions are instead characterized by large
neighborhoods.
To give quantitative references, we find that for $50\%$ of the networks,
 $\left| N^{2}_{\max} \right| \geq 0.22 \, n$.

In summary, these results show that some networks contain large neighborhoods.
If the NMP equations scale poorly with the neighborhood size then the
approach will be infeasible.
In the next section, we remedy this by adjusting $r$ at the level of individual nodes.

\section{Heterogeneous message passing}
\label{sec:het_NMP}

\subsection{General approach}

In the NMP approach of~\cite{cantwell2019message} one chooses a single value of $r$.
Using the neighborhoods, and specific to the problem at hand, one derives a set of  MP equations of the form
\change{
\begin{align}
	H_i &= \Phi_i \big( \bm{H}_{i \leftarrow N_i^r} \big) \\
	H_{i \leftarrow j} &= \Phi_{i \leftarrow j} \big( \bm{H}_{i
                             \leftarrow N_j^r \setminus N_i^r} \big)
                             \; ,
\end{align}
where $\bm{H}_{i \leftarrow N_i^r}$ is the vector of $H_{i \leftarrow
  j}$ for $j \in N_i^r$, and $\Phi_i$ and $\Phi_{i \leftarrow j}$ are problem-specific MP functions.}
A MP algorithm consists of \change{initializing the variables (e.g., at random) and then }
iterating this set of equations until they converge to a fixed point.
\change{In general, convergence may not be mathematically guaranteed; necessary and sufficient conditions for convergence are unclear, but this does not appear to be a significant problem in practice \cite{weiss2000correctness, zivan2020beyond}.}
When setting
$r=0$,
these equations reduce to the conventional MP ones.
Increasing $r$ should increase the accuracy of the approach, but
potentially
with a considerable increase of the computational cost.

As discussed, in a heterogeneous network there may be competing
considerations on how to appropriately set the value of $r$.
For example, in a sparse network with a high density of short cycles, we are likely to require $r \geq 1$ for an accurate approximation.
On the other hand, heterogeneous degree distributions may even increase the cost to solving the traditional MP equations.
Further increasing the complexity of the equations to be solved by increasing $r$ may simply be untenable.
How should one proceed?

Our solution is based on the simple observation that the neighborhood formalism does not actually require that each neighborhood is constructed with the same value of $r$.
Around nodes that are dense with short cycles but have relatively low degree, we can increase $r$. 
This increases accuracy with only a small increase to computational cost.
Conversely, for nodes with very high degrees, we can decrease $r$ to
reduce computational burden without a significant decrease in accuracy.
In fact, for nodes with extremely high degree we should find \change{$H_{i \leftarrow j} \approx H_{j}$}, i.e., that the messages have the same numerical value as the marginals.
Making this approximation corresponds to a mean-field approximation, and helps to further reduce the computational cost caused by high-degree nodes.

We allow each node $i$ to have its own approximation value $r_i$.
Re-writing the NMP equations but with heterogeneous $r$ we get
\change{
\begin{align}\label{eq:hetNMP}
	H_i &= \Phi_i \big( \bm{H}_{i \leftarrow N_i^{r_i}} \big) \\
	H_{i \leftarrow j} &= 
	\begin{cases}
		\, H_{j} \quad &\text{if } r_j=-1\\
		\, \Phi_{i \leftarrow j} \big( \bm{H}_{i \leftarrow N_{j}^{r_j} \setminus N_i^{r_i} } \big) \quad &\text{if } r_j \geq 0.
	\end{cases}\label{eq:message}
\end{align}
}
Note we allow $r_j=-1$ and use this notation to indicate the standard mean-field approximation.

Below, we test the ability of heterogeneous NMP to account for the spectral properties of large networks.
We find that increasing $r$ does indeed increase the accuracy of the approach over conventional MP\change{, at the expense of increased compute time}.
\change{However, by} setting $r_i=0$ for large-degree nodes, we retain much of the improved accuracy for only a small additional cost compared to conventional MP.
Remarkably, by setting $r_i=-1$ for the large-degree nodes, we find it is possible to \change{derive an algorithm that is able to improve on} both accuracy and speed, \change{compared to conventional MP}.

\subsection{Spectral density estimation}

As a specific application, we consider an heterogeneous NMP
approximation for the estimation of the spectral density of the graph
operators, e.g., adjacency matrix, graph Laplacian.
The spectral density of matrix $\bm{M}$ with eigenvalues $\lambda_k$ is defined
\begin{align}
	\rho(z) = - {1 \over n \pi} \sum_{k=1}^{n} {1 \over z - \lambda_k}
\end{align}
for complex $z$.

Following Ref.~\cite{cantwell2019message}, one can approximate $\rho(z)$ by first solving the MP equations
\begin{equation}
	H_{i} (z) = \sum_{w \in W_i} \vert w \vert \prod_{j \in w} {1 \over z - H_{i \leftarrow j}(z)}
\end{equation}
\begin{equation}
	H_{i \leftarrow j} (z) = 
		\sum_{w \in W_{j\setminus i}} \vert w \vert \prod_{k
                  \in w} {1 \over z - H_{j \leftarrow k}(z)}  \; ,
\end{equation}
where the sum is over all closed walks $w$ in the neighborhood $N^0_i$ or $N^0_j \setminus N^0_i$ respectively, and $\vert w \vert$ is the product of all edges in the walk.
These equations can be solved relatively efficiently using matrix
algebra---see \cite{cantwell2019message}---and finally one approximates
\begin{equation}
	\rho(z) = - {1 \over n \pi} \sum_{i=1}^{n} {1 \over z -
          H_i(z)} \; .
      \end{equation}

In the NMP heterogeneous approximation,  we mostly leave the equations
unchanged, except that now we
allow the neighborhood $N^{r_i}_i$ of node $i$ to be defined
with its own value of $r_i$, and also allow for
the mean-field approximation if $r_i=-1$,
\begin{equation}
	H_{i \leftarrow j} (z) = 
	\begin{cases}
		H_{j}(z) \quad &r_j=-1 \\
		{\sum_{w \in W_{j \setminus i}}} \vert w \vert {\prod_{k \in w}} {1 \over z - H_{j \leftarrow k}(z)} \quad &r_j \geq 0.
	\end{cases}
\end{equation}

To establish the desired value of $r_i$ for each node~$i$,
we set parameters $r_{\text{min}}$, $r_{\text{max}}$, and $K$.
The value of $r_i$ is chosen to be the largest value $r_{\text{min}} \leq r_i \leq r_{\text{max}}$ so that $\big\vert N_i^{r_i} \big\vert \leq K$.
This procedure imposes $r_i = r_{\text{min}}$ whenever the degree 
of node~$i$ is larger than $K$.
Otherwise, it imposes $r_i$ as large as possible so that the
neighborhood contains no more than $K$ nodes.

\section{Numerical results}
\label{sec:results}

We estimate the spectral density of the graph Laplacian of the real
networks in our corpus.
\change{The spectral properties of the Laplacian are important for 
many graph applications, including graph invariants (e.g., connectivity, expanding properties, genus, diameter, mean distance and chromatic number), partition problems (e.g, graph bisection,
connectivity and separation, isoperimetric numbers, maximum cut,
clustering, graph partition), and optimization
problems (e.g., cutwidth, bandwidth, min-p-sum problems,
ranking, scaling, quadratic assignment problem)~\cite{merris94, chung03, chung97, biyikoglu07}.}
\change{We estimate the} ground-truth density,
namely $\rho(z)$, using
the standard numerical library LAPACK~\cite{lapack}.
This method requires a time that scales as the cube of
the network size.

NMP approximations, denoted with $\tilde{\rho}(z)$,  are instead
obtained  
%using the above-described procedure, 
\change{numerically solving Eqs.~(9-10)}. We consider different levels of approximations. We always set
$r_{\min} = -1$; we consider $r_{\max} =0$, $r_{\max} =1$ and
$r_{\max}=2$; we vary the value of the parameter $K$.
For $K = N \geq | N_{\max}^r|$, no heterogeneous approximation is {\it de facto}
implemented, and the above approach reduces to the one already
considered in Ref.~\cite{cantwell2019message}. 

The densities $\rho(z)$
and $\tilde{\rho}(z)$ are computed
for $z \in [0, 10]$ and with a resolution $dz = 0.2$. In particular,
we normalize the densities within the interval $[0, 10]$.
In the NMP approximations, we use $\epsilon = 0.1$ as the value of the broadening
parameter, see Ref. ~\cite{cantwell2019message} for details.
All numerical tests are performed on
a server with Intel(R) Xeon(R) CPU E5-2690 v4 @ 2.60GHz CPUs and $378$
GB of RAM.

Not all the networks in our corpus are part of the analysis.
For $K=N$, we consider only networks with a number of
nodes $N \leq 20,000$.
For $K = 10$, we consider all networks with size $N \leq 100,000$.
Irrespective of the level of the approximation,
we let the algorithm run for up to seven days on our machine.
For the slowest NMP approximation, i.e., $r_{\max}=2$ and $K=N$,
we were able to estimate the spectral density of the graph Laplacian
only for $41$ networks. For faster approximations,
the number of analyzed networks was higher.
Details are provided in Tables~\ref{tab:1}-~\ref{tab:3}.

We first focus our attention on how the
time required for the estimation of the spectral density of the graph
Laplacian using the NMP approximations scales with size of the network
and the size of the largest neighborhood in the network.

\begin{figure}[!htb]
  \includegraphics[width=0.5\textwidth]{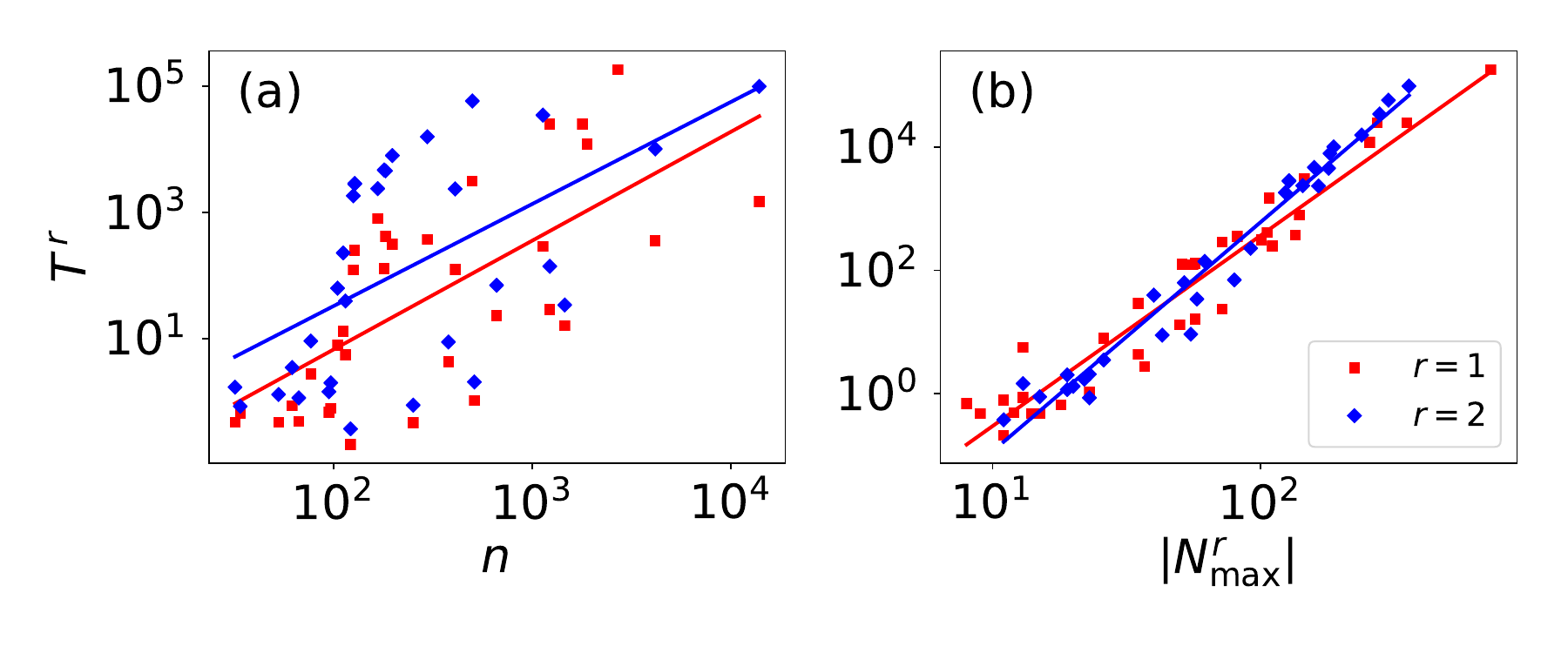}
  \caption{
    {\bf Computational time for the estimation of the spectral density
      of the graph Laplacian. } 
    We consider a subset of the $109$ real-world networks at our
    disposal. For each network, we
measure the computational time $T^r$ required by the NMP
algorithm for the computation of the spectral density of the graph
Laplacian. Estimates of $T^r$ are given in seconds.
{\bf (a)} We plot $T^r$ as a function of the network size $n$.
The lines are power-law fits of the data points, i.e., $T^r
\sim n^{\tau^r}$. Fits are obtained using simple linear regression
between the log-transformed variables.
We find $\tau^1 = 1.21$ (Pearson linear correlation coefficient $0.64$) and
$\tau^2 = 1.32 (0.48)$.
	{\bf (b)} We consider the same networks as in panel (a), but
we test the scaling $T^r
\sim |N_{\text{max}}^{r}|^{\sigma^r}$. We find $\sigma^1 = 2.99 (0.96)$ and
$\sigma^2 = 3.56 (0.98)$. 
}
    \label{fig:3}
  \end{figure}

Results for $K=N$ are
presented in Fig.~\ref{fig:3}. The relation between computational time
$T^{r}$ and network size $n$ is not very neat (Fig.~\ref{fig:3}a).
However, $T^{r}$ grows power like with $\left|
  N^{r}_{\max} \right|$ in a clear manner (Fig.~\ref{fig:3}b). The
measured exponents are all in line with the expected
complexity of the matrix inversion algorithm used to estimate messages
within individual neighborhoods~\cite{cantwell2019message}. In fact, the
inversion algorithm scales cubically with the matrix dimension, thus
the computational time of the entire algorithm is dominated by the
inversion of the matrix associated with the largest neighborhood in
the graph.

\begin{figure}[!htb]
  \includegraphics[width=0.5\textwidth]{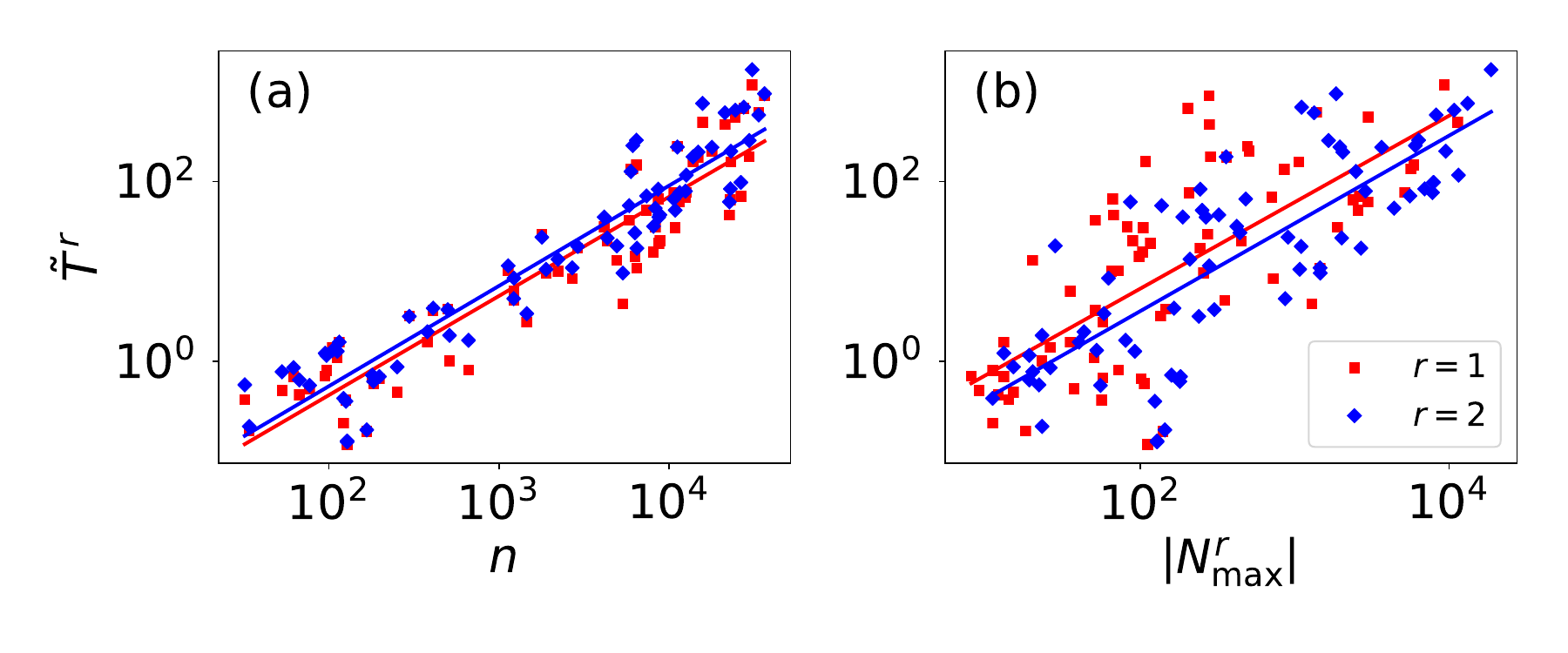}
  \caption{
    {\bf Computational time for the estimation of the spectral density
      of the graph Laplacian under the heterogeneous approximation. }
     We consider a subset of the $109$ real-world networks at our
    disposal.  For
    each network, we
measure the computational time $\tilde{T}^r$ required by the NMP
algorithm for the computation of the spectral density of the graph
Laplacian. Estimates of $\tilde{T}^r$ are given in seconds.
{\bf (a)} We plot $\tilde{T}^r$ as a function of the network size $n$.
The lines are power-law fits of the data points, i.e., $\tilde{T}^r
\sim n^{\tilde{\tau}^r}$. We find $\tilde{\tau}^1 = 0.98$  (Pearson linear correlation coefficient $0.93$) and
	$\tilde{\tau}^2 = 1.00 (0.94)$. {\bf (b)} We consider the same networks as in panel (a), but
we test the scaling $\tilde{T}^r
\sim |N_{\text{max}}^{r}|^{\tilde{\sigma}^r}$. We find $\tilde{\sigma}^1 = 0.80 (0.66)$ and
$\tilde{\sigma}^2 = 0.81 (0.72)$.}
    \label{fig:4}
  \end{figure}

Results for $K=10$ are
presented in Fig.~\ref{fig:4}. The relation between computational time
$\tilde{T}^{r}$ and network size $n$ is clearly linear (Fig.~\ref{fig:4}a).
$\tilde{T}^{r}$ now grows sub-linearly with $\left|
  N^{r}_{\max} \right|$, however, the relationship is not
as clear as the one that relates $\tilde{T}^{r}$ to $n$ (Fig.~\ref{fig:4}b).

\begin{figure}[!htb]
  \includegraphics[width=0.5\textwidth]{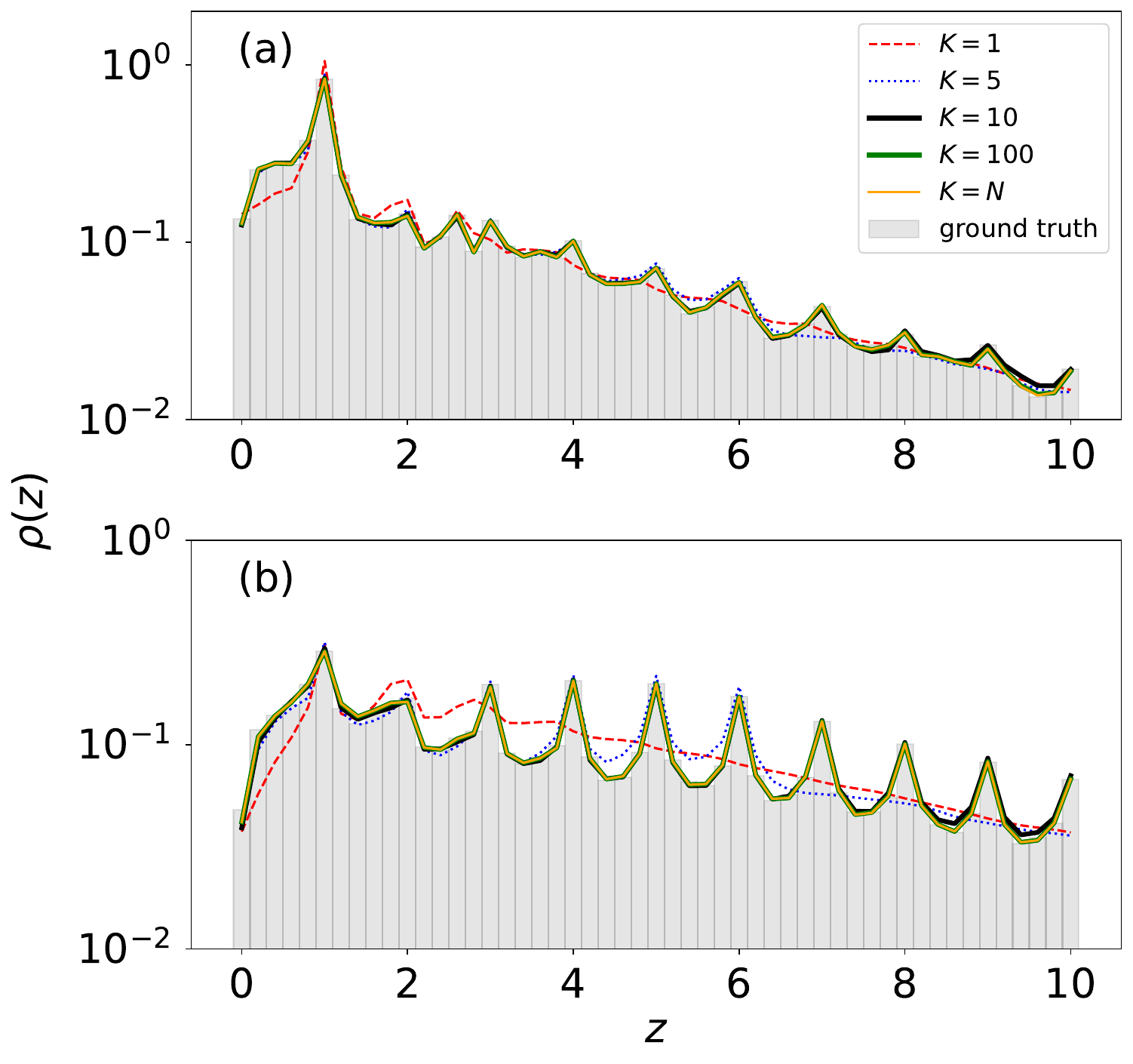}
  \caption{
    {\bf Laplacian spectral density of real-world networks.}
    {\bf (a)} Spectral density of the graph
    Laplacian $\rho(z)$ for the network of users of the
    Pretty-Good-Privacy algorithm for secure information
    interchange~\cite{boguna2004models}. The various approximations
    are obtained by setting
     $r_{\min}=-1$, $r_{\max}=2$, but different $K$ values.
    The ground-truth
    density is estimated using LAPACK.  {\bf (b)} Same
	 as in panel (a), but for the Cond-Mat collaboration
    network~\cite{newman2001structure}.
  }
    \label{fig:5}
  \end{figure}

 In Fig.~\ref{fig:5}, we compare the ground-truth spectral density 
 with NMP-based estimates obtained at different levels of
 approximation.
 Here, we set $r_{\min}=-1$ and $r_{\max}=2$, and we vary $K$
 to control for the level of the approximation.
 The comparison is made for two real-world networks. For small $K$
 values, the approximate spectral density fails to properly capture
 the behavior of the ground-truth density.
 The accuracy greatly improves as $K$ is increased if $K$ is
 small. For sufficiently large values of $K$, no visible changes are
 apparent in the estimated densities. For example, 
 already for $K=10$ the approximate density appears almost identical
 to the one
 obtained for $K=N$.

  \begin{figure}[!htb]
  \includegraphics[width=0.5\textwidth]{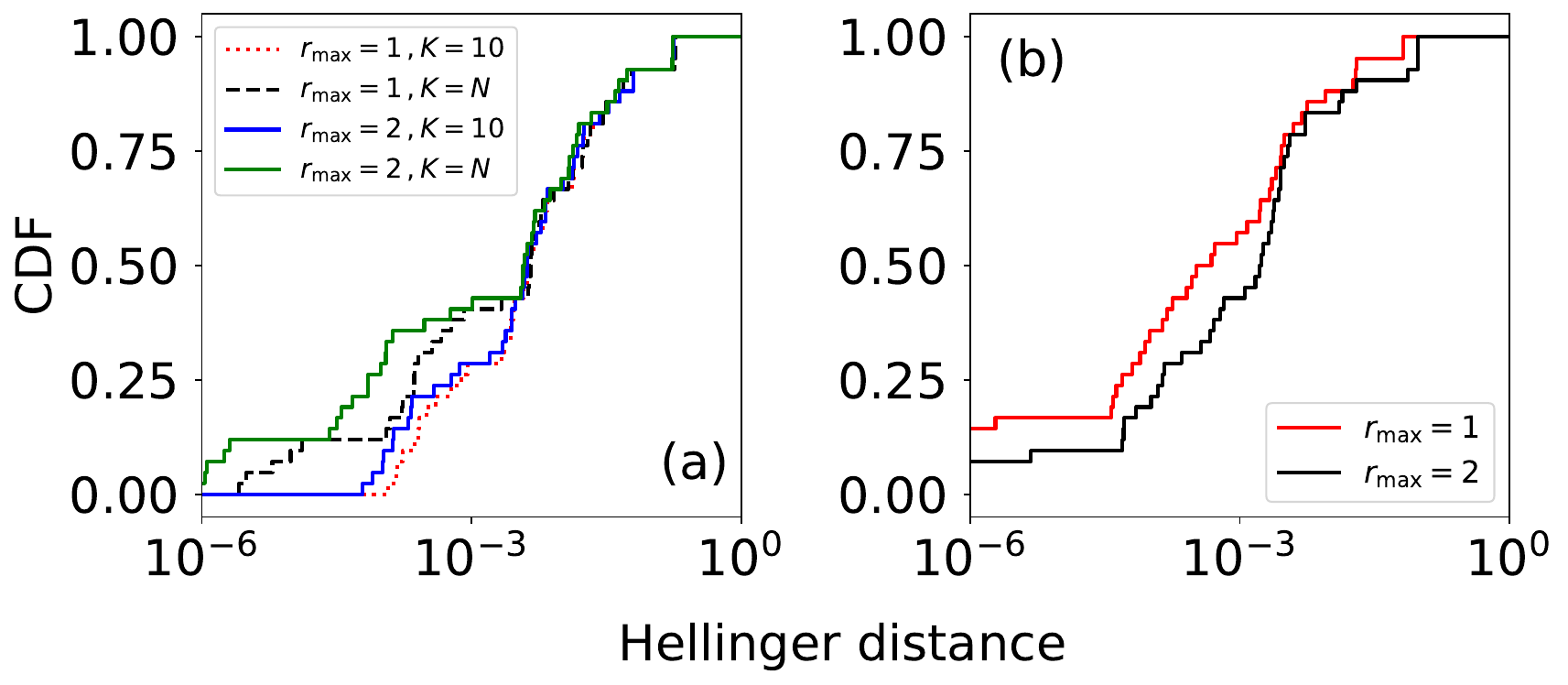}
  \caption{
    {\bf Accuracy of NMP approximations in
      reproducing the Laplacian spectral density of real-world networks.}
    {\bf (a)} We consider a subset of the $109$ real-world networks at our
    disposal.  For
    each network, we estimate the Hellinger distance
    between the
    ground-truth Laplacian spectral $\rho(z)$ and its approximation
    $\tilde{\rho}(z)$
    obtained via NMP. In all approximations, we set $r_{\min}=-1$. We
    vary instead the values of $r_{\max}$ and $K$. For each
    approximation, we plot the
    cumulative distribution function (CDF) of the Hellinger distance
    over the set of analyzed networks.  {\bf (b)}  For each
  network, we estimate the Hellinger distance between the NMP
  approximations obtained for $K=10$ and $K=N$. Results of the
  experiments are obtained for $r_{\min}=-1$. We consider 
  the cases $r_{\max} = 1$ and $r_{\max} = 2$.}
    \label{fig:6}
  \end{figure}

 We test systematically the above two observations in the corpus of real
 networks. To compare two spectral densities, we 
 make use of
the Hellinger distance, i.e., 
\change{
\begin{equation}
d (\rho, \tilde{\rho}) = 1 -
\int_{0}^{10} \, dz \sqrt{\rho(z) \tilde{ \rho}(z) }. 
\end{equation}
By definition, we have $0 \leq d (\rho, \tilde{\rho}) \leq 1$, with $d=0$ indicating perfect agreement between $\rho$ and $\tilde{\rho}$.}
 In Fig.~\ref{fig:6}a, we
 display the cumulative distribution of the  Hellinger distance
 obtained over a subset of networks in our corpus. Comparisons are
 made between the ground-truth density and different types of
 approximations. All NMP-based approximations are generally good.
 The accuracy of the approximation increases if we
 increase $r_{\max}=1$ to $r_{\max}=2$, and also we increase $K=10$ to
 $K=N$. However, the change in accuracy is not that dramatic.
 Indeed, for a fixed value of $r_{\max}$, increasing $K=10$ to $K=N$ generates
 little changes in the predicted distribution as apparent from the
 results of Fig.~\ref{fig:6}b. For $75\%$ of the networks, the
 increase $K=10 \to K=N$ induces a change in the predicted distribution
 corresponding to a value of the
Hellinger distance
 smaller than $0.01$.

  \begin{figure}[!htb]
  \includegraphics[width=0.5\textwidth]{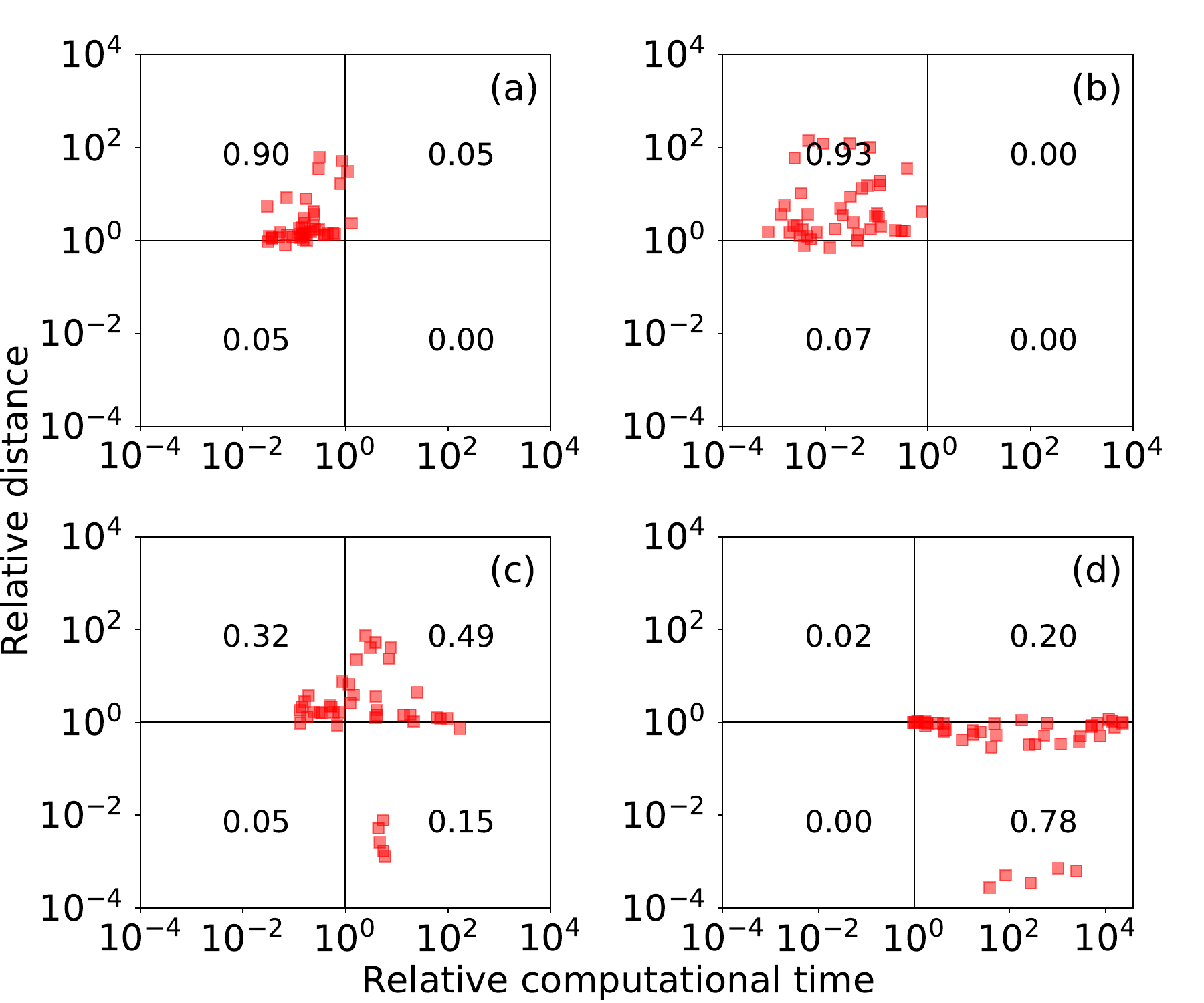}
  \caption{
    {\bf Trade off between time and accuracy in approximating
      the Laplacian spectral density of real-world networks.}
    {\bf (a)} We consider the same subset of real-networks as in
    Fig.~\ref{fig:6}. Each point denotes a network. For each
    network, we estimate the Hellinger distances
    between the  ground-truth Laplacian spectral density and two NMP
    approximations, namely approximations $a$ and $b$,
    obtained for $ r_{\max} = r^{(a)}_{\max} = 0$, $K
    = K^{(a)}=N$
    and $r_{\max}  = r^{(b)}_{\max} = 1$, $K
    = K^{(b)}=N$, respectively. We also measure the computation time
    required by the approximations. We then plot 
    %the logarithm of 
    the
    ratio between the Hellinger distances of two approximations
    as a function of 
    %the logarithm of
    the ratio of their computational time.
    To facilitate the interpretation of the plot, we divided it into
    four quadrants. The left quadrants denote
   the region where approximation $a$ is slower than approximation
   $b$; the top quadrants indicate that approximation $a$ is
   less accurate than approximation $b$.
   We also report the fraction of networks observed in each of the
   four quadrants. {\bf (b)} Same as
   in (a), but for $r^{(a)}_{\max} = 0$, $K
    = K^{(a)}=N$ and $r^{(b)}_{\max} = 2$ and $K
    = K^{(b)}=N$.  {\bf (c)} Same as
   in (a), but for $r^{(a)}_{\max} = 0$, $K
    = K^{(a)}=N$ and $r^{(b)}_{\max} = 2$ and $K
    = K^{(b)}=10$.  {\bf (d)} Same as
   in (a), but for $r^{(a)}_{\max} = 2$, $K
    = K^{(a)}=N$ and $r^{(b)}_{\max} = 2$ and $K
    = K^{(b)}=10$. 
  }
    \label{fig:7}
  \end{figure}

We complete the summary of our analysis in Fig.~\ref{fig:7}. There, we
compare different levels of approximations in terms of accuracy and
computational time. As expected, we find that increasing $r_{\max}$
while keeping $K$ fixed leads to an increase of accuracy
and computational time (Figs.~\ref{fig:7}a and ~\ref{fig:7}b).
Only a few exceptions are visible; these are given by very small
networks, with sizes $n \leq 100$. Quite surprisingly, we find that,
for \change{$49\%$}
%about  $50\%$ 
of the analyzed networks,
the proposed heterogeneous NMP approximation is faster and
more accurate than the
standard MP
approximation
(Fig.~\ref{fig:7}c). For many small networks, their performance is
similar and is obtained in a similar time. The only apparent exceptions
are given by five distinct snapshots of the 
peer-to-peer Gnutella network~\cite{ripeanu2002mapping,
  leskovec2007graph}, where the
MP approximation
greatly outperforms the NMP approximation, but with a computational
time that is about two orders of magnitude larger than the NMP
approximation. Finally, we confirm that the accuracy of the NMP
approximation obtained for $r_{\max} =2$ and $K=10$ is almost
identical to the one achieved for $r_{\max} =2$ and $K=N$ (Fig.~\ref{fig:7}c). The only
exceptions are still given by the five Gnutella networks. However, the
greater accuracy is achieved thanks to a significant increase in
computational time.

\begin{figure}[!htb]
  \includegraphics[width=0.5\textwidth]{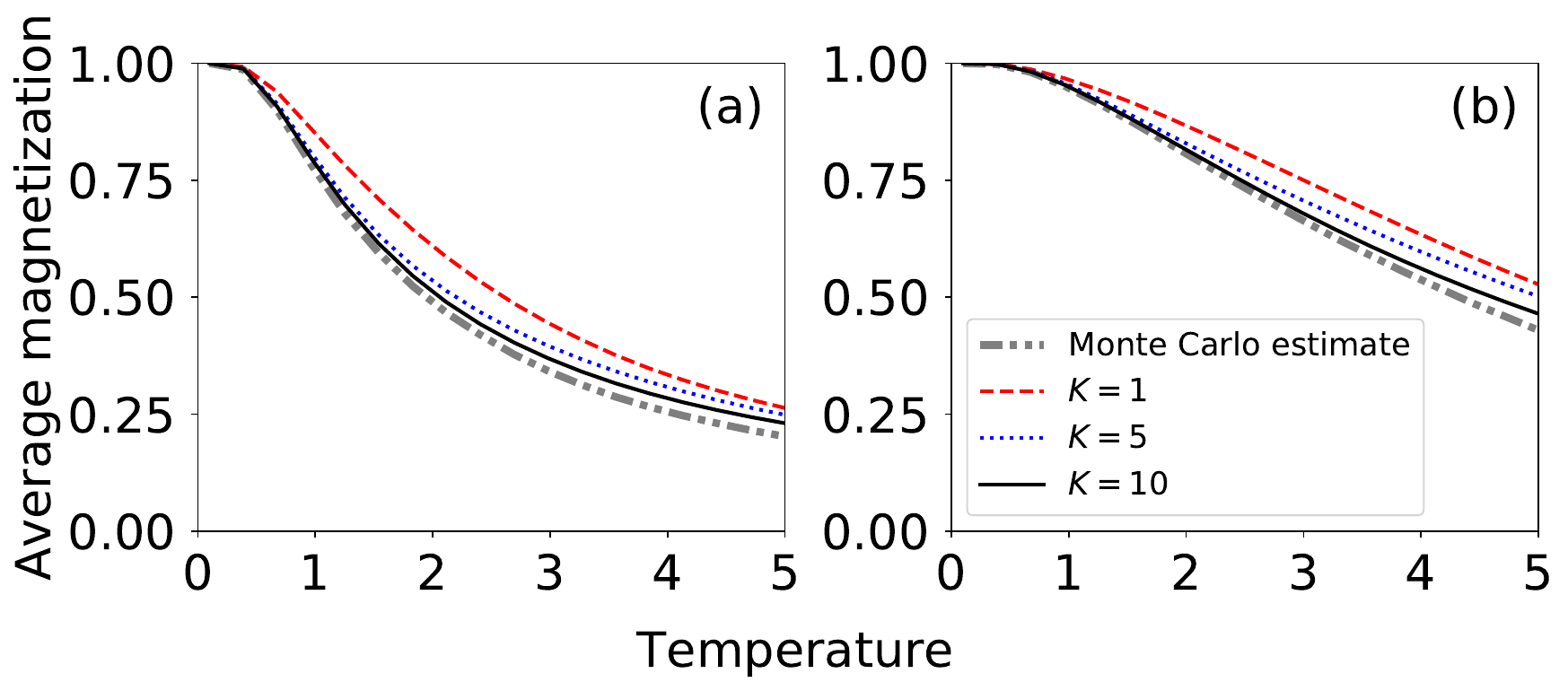}
  \caption{
    {\bf Heterogeneous belief propagation for the zero-field Ising model.}
    {\bf (a)} Average magnetization of the zero-field Ising model for
    the Pretty-Good-Privacy network~\cite{boguna2004models}. Results
    are obtained by integrating the heterogeneous NMP equations of
    Eqs.~(\ref{eq:hetNMP}) and~(\ref{eq:message}) with the belief
    propagation framework of~\cite{kirkley2021belief}. The various curves are obtained by setting
     $r_{\min}=-1$, $r_{\max}=2$, but allowing $K$ to range over multiple different values.
    The ground-truth magnetization is estimated using the Wolff cluster MCMC algorithm. 
    {\bf (b)} Same
	as in panel (a), but for the Cond-Mat collaboration
    network~\cite{newman2001structure}.
  }
    \label{fig:8}
  \end{figure}

\section{Discussion}
\label{sec:discussion}

We systematically tested the NMP approach on a corpus of $109$ networks.
We found, in accordance with expectations, that increasing the cycles
accounted for by increasing $r$ led to improved accuracy at the expense of speed.

We also introduced a hybrid approach, that uses more
accurate approximations for low-degree nodes,
and less accurate mean-field approximations at very high-degree nodes.
We tested the hybrid approach in estimating the spectral density of
the graph Laplacian of the real networks in our corpus.
Compared to conventional MP, this approach was more accurate in 81\% of networks \change{(Fig.~\ref{fig:7}(c): $0.49+0.32=0.81$)}.
In addition, it was also faster in 64\% of cases  \change{(Fig.~\ref{fig:7}(c): $0.49+0.15=0.64$)}.
In a plurality of cases, the approximation was both faster and more accurate than conventional MP.
The findings suggest that the NMP framework is applicable to a wide
class of networks,
even those with very large hubs.

The proposed NMP approach can be used in other problems whose solution can be approximated via conventional MP. 
Our detailed numerical experiments have focused on spectral density, a problem with ground truth that can be computed numerically for relatively large networks.
However, as a final example, we consider the belief propagation framework of~\cite{kirkley2021belief} to compute the magnetization
of the zero-field Ising model.
Fig.~\ref{fig:8} shows the
magnetization for $r_{\min}=-1$, $r_{\max}=2$ and different values of
$K$, for the same networks analyzed in Fig.~\ref{fig:5}. We see
that as we increase $K$ we get improved approximations of the
magnetization, and that for $K=10$ the results are quite close to 
%the ground truth.
\change{Monte Carlo estimates.}

In summary, we find that heterogeneous message passing approximations are effective for heterogeneous networks.
Specifically, high-degree nodes can be well accounted for with conventional mean-field approaches, while the corrections due to cycles are incorporated in the lower-degree nodes.
Both speed and accuracy can be simultaneously improved by applying appropriate approximations.

\acknowledgements{
  The authors thank Ginestra Bianconi, Cristopher Moore, and Mark
  Newman for helpful conversations.  The work was supported by
  National Science Foundation Grant BIGDATA-1838251 (G.T.C.), the Army Research Office Grant W911NF-21-1-0194 (F.R.), and the Air Force Office of Scientific Research Grant FA9550-21-1-0446 (F.R.).  The funders had no role in study design, data collection and analysis, decision to publish, or any opinions, findings, and conclusions or recommendations expressed in the paper. 
  }

%\bibliography{bibliography.bib}

\begin{thebibliography}{84}
\expandafter\ifx\csname natexlab\endcsname\relax\def\natexlab#1{#1}\fi
\expandafter\ifx\csname bibnamefont\endcsname\relax
  \def\bibnamefont#1{#1}\fi
\expandafter\ifx\csname bibfnamefont\endcsname\relax
  \def\bibfnamefont#1{#1}\fi
\expandafter\ifx\csname citenamefont\endcsname\relax
  \def\citenamefont#1{#1}\fi
\expandafter\ifx\csname url\endcsname\relax
  \def\url#1{\texttt{#1}}\fi
\expandafter\ifx\csname urlprefix\endcsname\relax\def\urlprefix{URL }\fi
\providecommand{\bibinfo}[2]{#2}
\providecommand{\eprint}[2][]{\url{#2}}

\bibitem[{\citenamefont{Newman}(2022)}]{newman2022message}
\bibinfo{author}{\bibfnamefont{M.~E.~J.} \bibnamefont{Newman}},
  \bibinfo{journal}{arXiv preprint arXiv:2211.05054}  (\bibinfo{year}{2022}).

\bibitem[{\citenamefont{Karrer and Newman}(2010)}]{karrer2010message}
\bibinfo{author}{\bibfnamefont{B.}~\bibnamefont{Karrer}} \bibnamefont{and}
  \bibinfo{author}{\bibfnamefont{M.~E.~J.} \bibnamefont{Newman}},
  \bibinfo{journal}{Physical Review E} \textbf{\bibinfo{volume}{82}},
  \bibinfo{pages}{016101} (\bibinfo{year}{2010}).

\bibitem[{\citenamefont{Lokhov et~al.}(2014)\citenamefont{Lokhov, M{\'e}zard,
  Ohta, and Zdeborov{\'a}}}]{lokhov2014inferring}
\bibinfo{author}{\bibfnamefont{A.~Y.} \bibnamefont{Lokhov}},
  \bibinfo{author}{\bibfnamefont{M.}~\bibnamefont{M{\'e}zard}},
  \bibinfo{author}{\bibfnamefont{H.}~\bibnamefont{Ohta}}, \bibnamefont{and}
  \bibinfo{author}{\bibfnamefont{L.}~\bibnamefont{Zdeborov{\'a}}},
  \bibinfo{journal}{Physical Review E} \textbf{\bibinfo{volume}{90}},
  \bibinfo{pages}{012801} (\bibinfo{year}{2014}).

\bibitem[{\citenamefont{Lokhov et~al.}(2015)\citenamefont{Lokhov, M{\'e}zard,
  and Zdeborov{\'a}}}]{lokhov2015dynamic}
\bibinfo{author}{\bibfnamefont{A.~Y.} \bibnamefont{Lokhov}},
  \bibinfo{author}{\bibfnamefont{M.}~\bibnamefont{M{\'e}zard}},
  \bibnamefont{and}
  \bibinfo{author}{\bibfnamefont{L.}~\bibnamefont{Zdeborov{\'a}}},
  \bibinfo{journal}{Physical Review E} \textbf{\bibinfo{volume}{91}},
  \bibinfo{pages}{012811} (\bibinfo{year}{2015}).

\bibitem[{\citenamefont{Altarelli et~al.}(2013)\citenamefont{Altarelli,
  Braunstein, Dall’Asta, and Zecchina}}]{altarelli2013optimizing}
\bibinfo{author}{\bibfnamefont{F.}~\bibnamefont{Altarelli}},
  \bibinfo{author}{\bibfnamefont{A.}~\bibnamefont{Braunstein}},
  \bibinfo{author}{\bibfnamefont{L.}~\bibnamefont{Dall’Asta}},
  \bibnamefont{and} \bibinfo{author}{\bibfnamefont{R.}~\bibnamefont{Zecchina}},
  \bibinfo{journal}{Journal of Statistical Mechanics: Theory and Experiment}
  \textbf{\bibinfo{volume}{2013}}, \bibinfo{pages}{P09011}
  (\bibinfo{year}{2013}).

\bibitem[{\citenamefont{Altarelli et~al.}(2014)\citenamefont{Altarelli,
  Braunstein, Dall’Asta, Wakeling, and Zecchina}}]{altarelli2014containing}
\bibinfo{author}{\bibfnamefont{F.}~\bibnamefont{Altarelli}},
  \bibinfo{author}{\bibfnamefont{A.}~\bibnamefont{Braunstein}},
  \bibinfo{author}{\bibfnamefont{L.}~\bibnamefont{Dall’Asta}},
  \bibinfo{author}{\bibfnamefont{J.~R.} \bibnamefont{Wakeling}},
  \bibnamefont{and} \bibinfo{author}{\bibfnamefont{R.}~\bibnamefont{Zecchina}},
  \bibinfo{journal}{Physical Review X} \textbf{\bibinfo{volume}{4}},
  \bibinfo{pages}{021024} (\bibinfo{year}{2014}).

\bibitem[{\citenamefont{Bianconi et~al.}(2021)\citenamefont{Bianconi, Sun,
  Rapisardi, and Arenas}}]{bianconi2021message}
\bibinfo{author}{\bibfnamefont{G.}~\bibnamefont{Bianconi}},
  \bibinfo{author}{\bibfnamefont{H.}~\bibnamefont{Sun}},
  \bibinfo{author}{\bibfnamefont{G.}~\bibnamefont{Rapisardi}},
  \bibnamefont{and} \bibinfo{author}{\bibfnamefont{A.}~\bibnamefont{Arenas}},
  \bibinfo{journal}{Physical Review Research} \textbf{\bibinfo{volume}{3}},
  \bibinfo{pages}{L012014} (\bibinfo{year}{2021}).

\bibitem[{\citenamefont{Decelle et~al.}(2011)\citenamefont{Decelle, Krzakala,
  Moore, and Zdeborov{\'a}}}]{decelle2011inference}
\bibinfo{author}{\bibfnamefont{A.}~\bibnamefont{Decelle}},
  \bibinfo{author}{\bibfnamefont{F.}~\bibnamefont{Krzakala}},
  \bibinfo{author}{\bibfnamefont{C.}~\bibnamefont{Moore}}, \bibnamefont{and}
  \bibinfo{author}{\bibfnamefont{L.}~\bibnamefont{Zdeborov{\'a}}},
  \bibinfo{journal}{Physical Review Letters} \textbf{\bibinfo{volume}{107}},
  \bibinfo{pages}{065701} (\bibinfo{year}{2011}).

\bibitem[{\citenamefont{Radicchi}(2018)}]{radicchi2018decoding}
\bibinfo{author}{\bibfnamefont{F.}~\bibnamefont{Radicchi}},
  \bibinfo{journal}{Physical Review E} \textbf{\bibinfo{volume}{97}},
  \bibinfo{pages}{022316} (\bibinfo{year}{2018}).

\bibitem[{\citenamefont{Radicchi and
  Castellano}(2018)}]{radicchi2018uncertainty}
\bibinfo{author}{\bibfnamefont{F.}~\bibnamefont{Radicchi}} \bibnamefont{and}
  \bibinfo{author}{\bibfnamefont{C.}~\bibnamefont{Castellano}},
  \bibinfo{journal}{Physical Review Letters} \textbf{\bibinfo{volume}{120}},
  \bibinfo{pages}{198301} (\bibinfo{year}{2018}).

\bibitem[{\citenamefont{Dorogovtsev et~al.}(2003)\citenamefont{Dorogovtsev,
  Goltsev, Mendes, and Samukhin}}]{dorogovtsev2003spectra}
\bibinfo{author}{\bibfnamefont{S.~N.} \bibnamefont{Dorogovtsev}},
  \bibinfo{author}{\bibfnamefont{A.~V.} \bibnamefont{Goltsev}},
  \bibinfo{author}{\bibfnamefont{J.~F.} \bibnamefont{Mendes}},
  \bibnamefont{and} \bibinfo{author}{\bibfnamefont{A.~N.}
  \bibnamefont{Samukhin}}, \bibinfo{journal}{Physical Review E}
  \textbf{\bibinfo{volume}{68}}, \bibinfo{pages}{046109}
  (\bibinfo{year}{2003}).

\bibitem[{\citenamefont{Rogers et~al.}(2008)\citenamefont{Rogers, Castillo,
  K{\"u}hn, and Takeda}}]{rogers2008cavity}
\bibinfo{author}{\bibfnamefont{T.}~\bibnamefont{Rogers}},
  \bibinfo{author}{\bibfnamefont{I.~P.} \bibnamefont{Castillo}},
  \bibinfo{author}{\bibfnamefont{R.}~\bibnamefont{K{\"u}hn}}, \bibnamefont{and}
  \bibinfo{author}{\bibfnamefont{K.}~\bibnamefont{Takeda}},
  \bibinfo{journal}{Physical Review E} \textbf{\bibinfo{volume}{78}},
  \bibinfo{pages}{031116} (\bibinfo{year}{2008}).

\bibitem[{\citenamefont{Karrer et~al.}(2014)\citenamefont{Karrer, Newman, and
  Zdeborov{\'a}}}]{karrer2014percolation}
\bibinfo{author}{\bibfnamefont{B.}~\bibnamefont{Karrer}},
  \bibinfo{author}{\bibfnamefont{M.~E.~J.} \bibnamefont{Newman}},
  \bibnamefont{and}
  \bibinfo{author}{\bibfnamefont{L.}~\bibnamefont{Zdeborov{\'a}}},
  \bibinfo{journal}{Physical Review Letters} \textbf{\bibinfo{volume}{113}},
  \bibinfo{pages}{208702} (\bibinfo{year}{2014}).

\bibitem[{\citenamefont{Hamilton and Pryadko}(2014)}]{hamilton2014tight}
\bibinfo{author}{\bibfnamefont{K.~E.} \bibnamefont{Hamilton}} \bibnamefont{and}
  \bibinfo{author}{\bibfnamefont{L.~P.} \bibnamefont{Pryadko}},
  \bibinfo{journal}{Physical Review Letters} \textbf{\bibinfo{volume}{113}},
  \bibinfo{pages}{208701} (\bibinfo{year}{2014}).

\bibitem[{\citenamefont{Radicchi and Castellano}(2015)}]{radicchi2015breaking}
\bibinfo{author}{\bibfnamefont{F.}~\bibnamefont{Radicchi}} \bibnamefont{and}
  \bibinfo{author}{\bibfnamefont{C.}~\bibnamefont{Castellano}},
  \bibinfo{journal}{Nature communications} \textbf{\bibinfo{volume}{6}},
  \bibinfo{pages}{1} (\bibinfo{year}{2015}).

\bibitem[{\citenamefont{Bianconi}(2017)}]{bianconi2017fluctuations}
\bibinfo{author}{\bibfnamefont{G.}~\bibnamefont{Bianconi}},
  \bibinfo{journal}{Physical Review E} \textbf{\bibinfo{volume}{96}},
  \bibinfo{pages}{012302} (\bibinfo{year}{2017}).

\bibitem[{\citenamefont{Bianconi}(2018)}]{bianconi2018rare}
\bibinfo{author}{\bibfnamefont{G.}~\bibnamefont{Bianconi}},
  \bibinfo{journal}{Physical Review E} \textbf{\bibinfo{volume}{97}},
  \bibinfo{pages}{022314} (\bibinfo{year}{2018}).

\bibitem[{\citenamefont{Mezard and Montanari}(2009)}]{mezard2009information}
\bibinfo{author}{\bibfnamefont{M.}~\bibnamefont{Mezard}} \bibnamefont{and}
  \bibinfo{author}{\bibfnamefont{A.}~\bibnamefont{Montanari}},
  \emph{\bibinfo{title}{Information, Physics, and Computation}}
  (\bibinfo{publisher}{Oxford University Press}, \bibinfo{year}{2009}).

\bibitem[{\citenamefont{MacKay}(2003)}]{mackay2003information}
\bibinfo{author}{\bibfnamefont{D.~J.~C.} \bibnamefont{MacKay}},
  \emph{\bibinfo{title}{Information Theory, Inference and Learning Algorithms}}
  (\bibinfo{publisher}{Cambridge University Press}, \bibinfo{year}{2003}).

\bibitem[{\citenamefont{Melnik et~al.}(2011)\citenamefont{Melnik, Hackett,
  Porter, Mucha, and Gleeson}}]{melnik2011unreasonable}
\bibinfo{author}{\bibfnamefont{S.}~\bibnamefont{Melnik}},
  \bibinfo{author}{\bibfnamefont{A.}~\bibnamefont{Hackett}},
  \bibinfo{author}{\bibfnamefont{M.~A.} \bibnamefont{Porter}},
  \bibinfo{author}{\bibfnamefont{P.~J.} \bibnamefont{Mucha}}, \bibnamefont{and}
  \bibinfo{author}{\bibfnamefont{J.~P.} \bibnamefont{Gleeson}},
  \bibinfo{journal}{Physical Review E} \textbf{\bibinfo{volume}{83}},
  \bibinfo{pages}{036112} (\bibinfo{year}{2011}).

\bibitem[{\citenamefont{Watts and Strogatz}(1998)}]{watts1998collective}
\bibinfo{author}{\bibfnamefont{D.~J.} \bibnamefont{Watts}} \bibnamefont{and}
  \bibinfo{author}{\bibfnamefont{S.~H.} \bibnamefont{Strogatz}},
  \bibinfo{journal}{nature} \textbf{\bibinfo{volume}{393}},
  \bibinfo{pages}{440} (\bibinfo{year}{1998}).

\bibitem[{\citenamefont{Newman}(2010)}]{newman2010networks}
\bibinfo{author}{\bibfnamefont{M.~E.~J.} \bibnamefont{Newman}},
  \emph{\bibinfo{title}{Networks}} (\bibinfo{publisher}{Oxford university
  press}, \bibinfo{year}{2010}).

\bibitem[{\citenamefont{Newman}(2009)}]{newman2009random}
\bibinfo{author}{\bibfnamefont{M.~E.~J.} \bibnamefont{Newman}},
  \bibinfo{journal}{Physical Review Letters} \textbf{\bibinfo{volume}{103}},
  \bibinfo{pages}{058701} (\bibinfo{year}{2009}).

\bibitem[{\citenamefont{Radicchi and Castellano}(2016)}]{radicchi2016beyond}
\bibinfo{author}{\bibfnamefont{F.}~\bibnamefont{Radicchi}} \bibnamefont{and}
  \bibinfo{author}{\bibfnamefont{C.}~\bibnamefont{Castellano}},
  \bibinfo{journal}{Physical Review E} \textbf{\bibinfo{volume}{93}},
  \bibinfo{pages}{030302} (\bibinfo{year}{2016}).

\bibitem[{\citenamefont{Cantwell and Newman}(2019)}]{cantwell2019message}
\bibinfo{author}{\bibfnamefont{G.~T.} \bibnamefont{Cantwell}} \bibnamefont{and}
  \bibinfo{author}{\bibfnamefont{M.~E.~J.} \bibnamefont{Newman}},
  \bibinfo{journal}{Proceedings of the National Academy of Sciences}
  \textbf{\bibinfo{volume}{116}}, \bibinfo{pages}{23398}
  (\bibinfo{year}{2019}).

\bibitem[{\citenamefont{Kirkley et~al.}(2021)\citenamefont{Kirkley, Cantwell,
  and Newman}}]{kirkley2021belief}
\bibinfo{author}{\bibfnamefont{A.}~\bibnamefont{Kirkley}},
  \bibinfo{author}{\bibfnamefont{G.~T.} \bibnamefont{Cantwell}},
  \bibnamefont{and} \bibinfo{author}{\bibfnamefont{M.~E.~J.}
  \bibnamefont{Newman}}, \bibinfo{journal}{Science Advances}
  \textbf{\bibinfo{volume}{7}}, \bibinfo{pages}{eabf1211}
  (\bibinfo{year}{2021}).

\bibitem[{\citenamefont{Barab{\'a}si and Albert}(1999)}]{barabasi1999emergence}
\bibinfo{author}{\bibfnamefont{A.-L.} \bibnamefont{Barab{\'a}si}}
  \bibnamefont{and} \bibinfo{author}{\bibfnamefont{R.}~\bibnamefont{Albert}},
  \bibinfo{journal}{science} \textbf{\bibinfo{volume}{286}},
  \bibinfo{pages}{509} (\bibinfo{year}{1999}).

\bibitem[{\citenamefont{Bianconi and Marsili}(2005)}]{bianconi2005loops}
\bibinfo{author}{\bibfnamefont{G.}~\bibnamefont{Bianconi}} \bibnamefont{and}
  \bibinfo{author}{\bibfnamefont{M.}~\bibnamefont{Marsili}},
  \bibinfo{journal}{Journal of Statistical Mechanics: Theory and Experiment}
  \textbf{\bibinfo{volume}{2005}}, \bibinfo{pages}{P06005}
  (\bibinfo{year}{2005}).

\bibitem[{\citenamefont{Radicchi}(2015)}]{radicchi2015predicting}
\bibinfo{author}{\bibfnamefont{F.}~\bibnamefont{Radicchi}},
  \bibinfo{journal}{Physical Review E} \textbf{\bibinfo{volume}{91}},
  \bibinfo{pages}{010801} (\bibinfo{year}{2015}).

\bibitem[{\citenamefont{Weiss}(2000)}]{weiss2000correctness} 
\bibinfo{author}{\bibfnamefont{Y.}~\bibnamefont{Weiss}},
  \bibinfo{journal}{Neural Computation} \textbf{\bibinfo{volume}{12}},
  \bibinfo{pages}{1--41} (\bibinfo{year}{2000}).

\bibitem[{\citenamefont{Zivan et~al.}(2020)}]{zivan2020beyond} 
\bibinfo{author}{\bibfnamefont{R.}~\bibnamefont{Zivan}},
\bibinfo{author}{\bibfnamefont{O.}~\bibnamefont{Lev}},
\bibinfo{author}{\bibfnamefont{R.}~\bibnamefont{Galiki}},
  \bibinfo{journal}{Proceedings of the AAAI Conference on Artificial Intelligence} \textbf{\bibinfo{volume}{34}},
  \bibinfo{pages}{7333--7340} (\bibinfo{year}{2020}).


  \bibitem{merris94} R. Merris, Linear Algebra Appl. {\bf 197–198},
143–176 (1994).

\bibitem{chung03} F. Chung, L. Lu and V. Vu, Proc. Natl. Acad. Sci. USA {\bf 100}, 6313–6318 (2003).

\bibitem{chung97} F. Chung, Spectral Graph Theory (CBMS Regional Conference Series in Mathematics, American Mathematical Society, 1997).

\bibitem{biyikoglu07}T. Biyikoglu, T. Leydold and  P. F. Stadler, Laplacian Eigenvectors of Graphs: Perron–Frobenius and Faber–Krahn Type Theorems (Lecture Notes in
Mathematics, Springer, 2007).

\bibitem[{lap()}]{lapack}
\emph{\bibinfo{title}{{LAPACK -- Linear Algebra PACKage}}},
  \bibinfo{howpublished}{\url{https://netlib.org/lapack/}},
  \bibinfo{note}{accessed: 2023-03-28}.

\bibitem[{\citenamefont{Bogu{\~n}{\'a}
  et~al.}(2004)\citenamefont{Bogu{\~n}{\'a}, Pastor-Satorras,
  D{\'\i}az-Guilera, and Arenas}}]{boguna2004models}
\bibinfo{author}{\bibfnamefont{M.}~\bibnamefont{Bogu{\~n}{\'a}}},
  \bibinfo{author}{\bibfnamefont{R.}~\bibnamefont{Pastor-Satorras}},
  \bibinfo{author}{\bibfnamefont{A.}~\bibnamefont{D{\'\i}az-Guilera}},
  \bibnamefont{and} \bibinfo{author}{\bibfnamefont{A.}~\bibnamefont{Arenas}},
  \bibinfo{journal}{Physical Review E} \textbf{\bibinfo{volume}{70}},
  \bibinfo{pages}{056122} (\bibinfo{year}{2004}).

\bibitem[{\citenamefont{Newman}(2001)}]{newman2001structure}
\bibinfo{author}{\bibfnamefont{M.~E.~J.} \bibnamefont{Newman}},
  \bibinfo{journal}{Proceedings of the National Academy of Sciences}
  \textbf{\bibinfo{volume}{98}}, \bibinfo{pages}{404} (\bibinfo{year}{2001}).

\bibitem[{\citenamefont{Ripeanu et~al.}(2002)\citenamefont{Ripeanu, Foster, and
  Iamnitchi}}]{ripeanu2002mapping}
\bibinfo{author}{\bibfnamefont{M.}~\bibnamefont{Ripeanu}},
  \bibinfo{author}{\bibfnamefont{I.}~\bibnamefont{Foster}}, \bibnamefont{and}
  \bibinfo{author}{\bibfnamefont{A.}~\bibnamefont{Iamnitchi}},
  \bibinfo{journal}{arXiv preprint cs/0209028}  (\bibinfo{year}{2002}).

\bibitem[{\citenamefont{Leskovec
  et~al.}(2007{\natexlab{a}})\citenamefont{Leskovec, Kleinberg, and
  Faloutsos}}]{leskovec2007graph}
\bibinfo{author}{\bibfnamefont{J.}~\bibnamefont{Leskovec}},
  \bibinfo{author}{\bibfnamefont{J.}~\bibnamefont{Kleinberg}},
  \bibnamefont{and}
  \bibinfo{author}{\bibfnamefont{C.}~\bibnamefont{Faloutsos}},
  \bibinfo{journal}{ACM Transactions on Knowledge Discovery from Data (TKDD)}
  \textbf{\bibinfo{volume}{1}}, \bibinfo{pages}{2}
  (\bibinfo{year}{2007}{\natexlab{a}}).

\bibitem[{\citenamefont{Milo et~al.}(2004)\citenamefont{Milo, Itzkovitz,
  Kashtan, Levitt, Shen-Orr, Ayzenshtat, Sheffer, and
  Alon}}]{milo2004superfamilies}
\bibinfo{author}{\bibfnamefont{R.}~\bibnamefont{Milo}},
  \bibinfo{author}{\bibfnamefont{S.}~\bibnamefont{Itzkovitz}},
  \bibinfo{author}{\bibfnamefont{N.}~\bibnamefont{Kashtan}},
  \bibinfo{author}{\bibfnamefont{R.}~\bibnamefont{Levitt}},
  \bibinfo{author}{\bibfnamefont{S.}~\bibnamefont{Shen-Orr}},
  \bibinfo{author}{\bibfnamefont{I.}~\bibnamefont{Ayzenshtat}},
  \bibinfo{author}{\bibfnamefont{M.}~\bibnamefont{Sheffer}}, \bibnamefont{and}
  \bibinfo{author}{\bibfnamefont{U.}~\bibnamefont{Alon}},
  \bibinfo{journal}{Science} \textbf{\bibinfo{volume}{303}},
  \bibinfo{pages}{1538} (\bibinfo{year}{2004}).

\bibitem[{\citenamefont{Zachary}(1977)}]{zachary1977information}
\bibinfo{author}{\bibfnamefont{W.~W.} \bibnamefont{Zachary}},
  \bibinfo{journal}{Journal of Anthropological Research} pp.
  \bibinfo{pages}{452--473} (\bibinfo{year}{1977}).

\bibitem[{\citenamefont{Lusseau et~al.}(2003)\citenamefont{Lusseau, Schneider,
  Boisseau, Haase, Slooten, and Dawson}}]{lusseau2003bottlenose}
\bibinfo{author}{\bibfnamefont{D.}~\bibnamefont{Lusseau}},
  \bibinfo{author}{\bibfnamefont{K.}~\bibnamefont{Schneider}},
  \bibinfo{author}{\bibfnamefont{O.~J.} \bibnamefont{Boisseau}},
  \bibinfo{author}{\bibfnamefont{P.}~\bibnamefont{Haase}},
  \bibinfo{author}{\bibfnamefont{E.}~\bibnamefont{Slooten}}, \bibnamefont{and}
  \bibinfo{author}{\bibfnamefont{S.~M.} \bibnamefont{Dawson}},
  \bibinfo{journal}{Behavioral Ecology and Sociobiology}
  \textbf{\bibinfo{volume}{54}}, \bibinfo{pages}{396} (\bibinfo{year}{2003}).

\bibitem[{\citenamefont{Knuth}(1993)}]{knuth1993stanford}
\bibinfo{author}{\bibfnamefont{D.~E.} \bibnamefont{Knuth}},
  \emph{\bibinfo{title}{The Stanford GraphBase: a platform for combinatorial
  computing}}, vol.~\bibinfo{volume}{37} (\bibinfo{publisher}{Addison-Wesley
  Reading}, \bibinfo{year}{1993}).

\bibitem[{\citenamefont{Mangan and Alon}(2003)}]{mangan2003structure}
\bibinfo{author}{\bibfnamefont{S.}~\bibnamefont{Mangan}} \bibnamefont{and}
  \bibinfo{author}{\bibfnamefont{U.}~\bibnamefont{Alon}},
  \bibinfo{journal}{Proceedings of the National Academy of Sciences}
  \textbf{\bibinfo{volume}{100}}, \bibinfo{pages}{11980}
  (\bibinfo{year}{2003}).

\bibitem[{\citenamefont{Adamic and Glance}(2005)}]{adamic2005political}
\bibinfo{author}{\bibfnamefont{L.~A.} \bibnamefont{Adamic}} \bibnamefont{and}
  \bibinfo{author}{\bibfnamefont{N.}~\bibnamefont{Glance}}, in
  \emph{\bibinfo{booktitle}{Proceedings of the 3rd international workshop on
  Link discovery}} (\bibinfo{organization}{ACM}, \bibinfo{year}{2005}), pp.
  \bibinfo{pages}{36--43}.

\bibitem[{\citenamefont{Newman}(2006)}]{newman2006finding}
\bibinfo{author}{\bibfnamefont{M.~E.~J.} \bibnamefont{Newman}},
  \bibinfo{journal}{Physical Review E} \textbf{\bibinfo{volume}{74}},
  \bibinfo{pages}{036104} (\bibinfo{year}{2006}).

\bibitem[{\citenamefont{Girvan and Newman}(2002)}]{girvan2002community}
\bibinfo{author}{\bibfnamefont{M.}~\bibnamefont{Girvan}} \bibnamefont{and}
  \bibinfo{author}{\bibfnamefont{M.~E.~J.} \bibnamefont{Newman}},
  \bibinfo{journal}{Proceedings of the National Academy of Sciences}
  \textbf{\bibinfo{volume}{99}}, \bibinfo{pages}{7821} (\bibinfo{year}{2002}).

\bibitem[{\citenamefont{Fournet and Barrat}(2014)}]{fournet2014contact}
\bibinfo{author}{\bibfnamefont{J.}~\bibnamefont{Fournet}} \bibnamefont{and}
  \bibinfo{author}{\bibfnamefont{A.}~\bibnamefont{Barrat}},
  \bibinfo{journal}{PloS one} \textbf{\bibinfo{volume}{9}},
  \bibinfo{pages}{e107878} (\bibinfo{year}{2014}).

\bibitem[{\citenamefont{Ulanowicz et~al.}(1998)\citenamefont{Ulanowicz,
  Bondavalli, and Egnotovich}}]{ulanowicz1998network}
\bibinfo{author}{\bibfnamefont{R.}~\bibnamefont{Ulanowicz}},
  \bibinfo{author}{\bibfnamefont{C.}~\bibnamefont{Bondavalli}},
  \bibnamefont{and}
  \bibinfo{author}{\bibfnamefont{M.}~\bibnamefont{Egnotovich}},
  \bibinfo{journal}{Annual Report to the United States Geological Service
  Biological Resources Division Ref. No.[UMCES] CBL} pp.
  \bibinfo{pages}{98--123} (\bibinfo{year}{1998}).

\bibitem[{\citenamefont{Kunegis}(2013)}]{konect}
\bibinfo{author}{\bibfnamefont{J.}~\bibnamefont{Kunegis}}, in
  \emph{\bibinfo{booktitle}{Proc. Int. Conf. on World Wide Web Companion}}
  (\bibinfo{year}{2013}), pp. \bibinfo{pages}{1343--1350},
  \urlprefix\url{http://userpages.uni-koblenz.de/~kunegis/paper/kunegis-koblenz-network-collection.pdf}.

\bibitem[{\citenamefont{Michalski et~al.}(2011)\citenamefont{Michalski, Palus,
  and Kazienko}}]{radoslaw}
\bibinfo{author}{\bibfnamefont{R.}~\bibnamefont{Michalski}},
  \bibinfo{author}{\bibfnamefont{S.}~\bibnamefont{Palus}}, \bibnamefont{and}
  \bibinfo{author}{\bibfnamefont{P.}~\bibnamefont{Kazienko}}, in
  \emph{\bibinfo{booktitle}{Lecture Notes in Business Information Processing}}
  (\bibinfo{publisher}{Springer Berlin Heidelberg}, \bibinfo{year}{2011}),
  vol.~\bibinfo{volume}{87}, pp. \bibinfo{pages}{197--206}.

\bibitem[{\citenamefont{Martinez}(1991)}]{martinez1991artifacts}
\bibinfo{author}{\bibfnamefont{N.~D.} \bibnamefont{Martinez}},
  \bibinfo{journal}{Ecological Monographs} pp. \bibinfo{pages}{367--392}
  (\bibinfo{year}{1991}).

\bibitem[{\citenamefont{Gleiser and Danon}(2003)}]{gleiser2003community}
\bibinfo{author}{\bibfnamefont{P.~M.} \bibnamefont{Gleiser}} \bibnamefont{and}
  \bibinfo{author}{\bibfnamefont{L.}~\bibnamefont{Danon}},
  \bibinfo{journal}{Advances in complex systems} \textbf{\bibinfo{volume}{6}},
  \bibinfo{pages}{565} (\bibinfo{year}{2003}).

\bibitem[{\citenamefont{Isella et~al.}(2011)\citenamefont{Isella, Stehl{\'e},
  Barrat, Cattuto, Pinton, and Van~den Broeck}}]{isella2011s}
\bibinfo{author}{\bibfnamefont{L.}~\bibnamefont{Isella}},
  \bibinfo{author}{\bibfnamefont{J.}~\bibnamefont{Stehl{\'e}}},
  \bibinfo{author}{\bibfnamefont{A.}~\bibnamefont{Barrat}},
  \bibinfo{author}{\bibfnamefont{C.}~\bibnamefont{Cattuto}},
  \bibinfo{author}{\bibfnamefont{J.-F.} \bibnamefont{Pinton}},
  \bibnamefont{and} \bibinfo{author}{\bibfnamefont{W.}~\bibnamefont{Van~den
  Broeck}}, \bibinfo{journal}{Journal of theoretical biology}
  \textbf{\bibinfo{volume}{271}}, \bibinfo{pages}{166} (\bibinfo{year}{2011}).

\bibitem[{\citenamefont{Colizza et~al.}(2007)\citenamefont{Colizza,
  Pastor-Satorras, and Vespignani}}]{colizza2007reaction}
\bibinfo{author}{\bibfnamefont{V.}~\bibnamefont{Colizza}},
  \bibinfo{author}{\bibfnamefont{R.}~\bibnamefont{Pastor-Satorras}},
  \bibnamefont{and}
  \bibinfo{author}{\bibfnamefont{A.}~\bibnamefont{Vespignani}},
  \bibinfo{journal}{Nature Physics} \textbf{\bibinfo{volume}{3}},
  \bibinfo{pages}{276} (\bibinfo{year}{2007}).

\bibitem[{\citenamefont{Milo et~al.}(2002)\citenamefont{Milo, Shen-Orr,
  Itzkovitz, Kashtan, Chklovskii, and Alon}}]{milo2002network}
\bibinfo{author}{\bibfnamefont{R.}~\bibnamefont{Milo}},
  \bibinfo{author}{\bibfnamefont{S.}~\bibnamefont{Shen-Orr}},
  \bibinfo{author}{\bibfnamefont{S.}~\bibnamefont{Itzkovitz}},
  \bibinfo{author}{\bibfnamefont{N.}~\bibnamefont{Kashtan}},
  \bibinfo{author}{\bibfnamefont{D.}~\bibnamefont{Chklovskii}},
  \bibnamefont{and} \bibinfo{author}{\bibfnamefont{U.}~\bibnamefont{Alon}},
  \bibinfo{journal}{Science} \textbf{\bibinfo{volume}{298}},
  \bibinfo{pages}{824} (\bibinfo{year}{2002}).

\bibitem[{\citenamefont{Guimera et~al.}(2003)\citenamefont{Guimera, Danon,
  Diaz-Guilera, Giralt, and Arenas}}]{guimera2003self}
\bibinfo{author}{\bibfnamefont{R.}~\bibnamefont{Guimera}},
  \bibinfo{author}{\bibfnamefont{L.}~\bibnamefont{Danon}},
  \bibinfo{author}{\bibfnamefont{A.}~\bibnamefont{Diaz-Guilera}},
  \bibinfo{author}{\bibfnamefont{F.}~\bibnamefont{Giralt}}, \bibnamefont{and}
  \bibinfo{author}{\bibfnamefont{A.}~\bibnamefont{Arenas}},
  \bibinfo{journal}{Physical Review E} \textbf{\bibinfo{volume}{68}},
  \bibinfo{pages}{065103} (\bibinfo{year}{2003}).

\bibitem[{\citenamefont{Jeong et~al.}(2001)\citenamefont{Jeong, Mason,
  Barab{\'a}si, and Oltvai}}]{jeong2001lethality}
\bibinfo{author}{\bibfnamefont{H.}~\bibnamefont{Jeong}},
  \bibinfo{author}{\bibfnamefont{S.~P.} \bibnamefont{Mason}},
  \bibinfo{author}{\bibfnamefont{A.-L.} \bibnamefont{Barab{\'a}si}},
  \bibnamefont{and} \bibinfo{author}{\bibfnamefont{Z.~N.}
  \bibnamefont{Oltvai}}, \bibinfo{journal}{Nature}
  \textbf{\bibinfo{volume}{411}}, \bibinfo{pages}{41} (\bibinfo{year}{2001}).

\bibitem[{\citenamefont{Opsahl and Panzarasa}(2009)}]{opsahl2009clustering}
\bibinfo{author}{\bibfnamefont{T.}~\bibnamefont{Opsahl}} \bibnamefont{and}
  \bibinfo{author}{\bibfnamefont{P.}~\bibnamefont{Panzarasa}},
  \bibinfo{journal}{Social networks} \textbf{\bibinfo{volume}{31}},
  \bibinfo{pages}{155} (\bibinfo{year}{2009}).

\bibitem[{\citenamefont{Bu et~al.}(2003)\citenamefont{Bu, Zhao, Cai, Xue, Zhu,
  Lu, Zhang, Sun, Ling, Zhang et~al.}}]{bu2003topological}
\bibinfo{author}{\bibfnamefont{D.}~\bibnamefont{Bu}},
  \bibinfo{author}{\bibfnamefont{Y.}~\bibnamefont{Zhao}},
  \bibinfo{author}{\bibfnamefont{L.}~\bibnamefont{Cai}},
  \bibinfo{author}{\bibfnamefont{H.}~\bibnamefont{Xue}},
  \bibinfo{author}{\bibfnamefont{X.}~\bibnamefont{Zhu}},
  \bibinfo{author}{\bibfnamefont{H.}~\bibnamefont{Lu}},
  \bibinfo{author}{\bibfnamefont{J.}~\bibnamefont{Zhang}},
  \bibinfo{author}{\bibfnamefont{S.}~\bibnamefont{Sun}},
  \bibinfo{author}{\bibfnamefont{L.}~\bibnamefont{Ling}},
  \bibinfo{author}{\bibfnamefont{N.}~\bibnamefont{Zhang}},
  \bibnamefont{et~al.}, \bibinfo{journal}{Nucleic acids research}
  \textbf{\bibinfo{volume}{31}}, \bibinfo{pages}{2443} (\bibinfo{year}{2003}).

\bibitem[{\citenamefont{Opsahl et~al.}(2010)\citenamefont{Opsahl, Agneessens,
  and Skvoretz}}]{opsahl2010node}
\bibinfo{author}{\bibfnamefont{T.}~\bibnamefont{Opsahl}},
  \bibinfo{author}{\bibfnamefont{F.}~\bibnamefont{Agneessens}},
  \bibnamefont{and} \bibinfo{author}{\bibfnamefont{J.}~\bibnamefont{Skvoretz}},
  \bibinfo{journal}{Social Networks} \textbf{\bibinfo{volume}{32}},
  \bibinfo{pages}{245} (\bibinfo{year}{2010}).

\bibitem[{\citenamefont{Radicchi}(2011)}]{radicchi2011best}
\bibinfo{author}{\bibfnamefont{F.}~\bibnamefont{Radicchi}},
  \bibinfo{journal}{PloS one} \textbf{\bibinfo{volume}{6}},
  \bibinfo{pages}{e17249} (\bibinfo{year}{2011}).

\bibitem[{\citenamefont{Joshi-Tope et~al.}(2005)\citenamefont{Joshi-Tope,
  Gillespie, Vastrik, D'Eustachio, Schmidt, de~Bono, Jassal, Gopinath, Wu,
  Matthews et~al.}}]{joshi2005reactome}
\bibinfo{author}{\bibfnamefont{G.}~\bibnamefont{Joshi-Tope}},
  \bibinfo{author}{\bibfnamefont{M.}~\bibnamefont{Gillespie}},
  \bibinfo{author}{\bibfnamefont{I.}~\bibnamefont{Vastrik}},
  \bibinfo{author}{\bibfnamefont{P.}~\bibnamefont{D'Eustachio}},
  \bibinfo{author}{\bibfnamefont{E.}~\bibnamefont{Schmidt}},
  \bibinfo{author}{\bibfnamefont{B.}~\bibnamefont{de~Bono}},
  \bibinfo{author}{\bibfnamefont{B.}~\bibnamefont{Jassal}},
  \bibinfo{author}{\bibfnamefont{G.}~\bibnamefont{Gopinath}},
  \bibinfo{author}{\bibfnamefont{G.}~\bibnamefont{Wu}},
  \bibinfo{author}{\bibfnamefont{L.}~\bibnamefont{Matthews}},
  \bibnamefont{et~al.}, \bibinfo{journal}{Nucleic acids research}
  \textbf{\bibinfo{volume}{33}}, \bibinfo{pages}{D428} (\bibinfo{year}{2005}).

\bibitem[{\citenamefont{{\v{S}}ubelj and Bajec}(2012)}]{vsubelj2012software}
\bibinfo{author}{\bibfnamefont{L.}~\bibnamefont{{\v{S}}ubelj}}
  \bibnamefont{and} \bibinfo{author}{\bibfnamefont{M.}~\bibnamefont{Bajec}}, in
  \emph{\bibinfo{booktitle}{Proceedings of the First International Workshop on
  Software Mining}} (\bibinfo{organization}{ACM}, \bibinfo{year}{2012}), pp.
  \bibinfo{pages}{9--16}.

\bibitem[{\citenamefont{Leskovec et~al.}(2005)\citenamefont{Leskovec,
  Kleinberg, and Faloutsos}}]{leskovec2005graphs}
\bibinfo{author}{\bibfnamefont{J.}~\bibnamefont{Leskovec}},
  \bibinfo{author}{\bibfnamefont{J.}~\bibnamefont{Kleinberg}},
  \bibnamefont{and}
  \bibinfo{author}{\bibfnamefont{C.}~\bibnamefont{Faloutsos}}, in
  \emph{\bibinfo{booktitle}{Proceedings of the eleventh ACM SIGKDD
  international conference on Knowledge discovery in data mining}}
  (\bibinfo{organization}{ACM}, \bibinfo{year}{2005}), pp.
  \bibinfo{pages}{177--187}.

\bibitem[{\citenamefont{Ley}(2002)}]{ley2002dblp}
\bibinfo{author}{\bibfnamefont{M.}~\bibnamefont{Ley}}, in
  \emph{\bibinfo{booktitle}{String Processing and Information Retrieval}}
  (\bibinfo{organization}{Springer}, \bibinfo{year}{2002}), pp.
  \bibinfo{pages}{1--10}.

\bibitem[{\citenamefont{Palla et~al.}(2007)\citenamefont{Palla, Farkas,
  Pollner, Derenyi, and Vicsek}}]{palla2007directed}
\bibinfo{author}{\bibfnamefont{G.}~\bibnamefont{Palla}},
  \bibinfo{author}{\bibfnamefont{I.~J.} \bibnamefont{Farkas}},
  \bibinfo{author}{\bibfnamefont{P.}~\bibnamefont{Pollner}},
  \bibinfo{author}{\bibfnamefont{I.}~\bibnamefont{Derenyi}}, \bibnamefont{and}
  \bibinfo{author}{\bibfnamefont{T.}~\bibnamefont{Vicsek}},
  \bibinfo{journal}{New Journal of Physics} \textbf{\bibinfo{volume}{9}},
  \bibinfo{pages}{186} (\bibinfo{year}{2007}).

\bibitem[{\citenamefont{Kiss et~al.}(1973)\citenamefont{Kiss, Armstrong,
  Milroy, and Piper}}]{kiss1973associative}
\bibinfo{author}{\bibfnamefont{G.~R.} \bibnamefont{Kiss}},
  \bibinfo{author}{\bibfnamefont{C.}~\bibnamefont{Armstrong}},
  \bibinfo{author}{\bibfnamefont{R.}~\bibnamefont{Milroy}}, \bibnamefont{and}
  \bibinfo{author}{\bibfnamefont{J.}~\bibnamefont{Piper}},
  \bibinfo{journal}{The computer and literary studies} pp.
  \bibinfo{pages}{153--165} (\bibinfo{year}{1973}).

\bibitem[{\citenamefont{{\v{S}}ubelj and Bajec}(2013)}]{vsubelj2013model}
\bibinfo{author}{\bibfnamefont{L.}~\bibnamefont{{\v{S}}ubelj}}
  \bibnamefont{and} \bibinfo{author}{\bibfnamefont{M.}~\bibnamefont{Bajec}}, in
  \emph{\bibinfo{booktitle}{Proceedings of the 22nd international conference on
  World Wide Web companion}} (\bibinfo{organization}{International World Wide
  Web Conferences Steering Committee}, \bibinfo{year}{2013}), pp.
  \bibinfo{pages}{527--530}.

\bibitem[{\citenamefont{De~Choudhury et~al.}(2009)\citenamefont{De~Choudhury,
  Sundaram, John, and Seligmann}}]{de2009social}
\bibinfo{author}{\bibfnamefont{M.}~\bibnamefont{De~Choudhury}},
  \bibinfo{author}{\bibfnamefont{H.}~\bibnamefont{Sundaram}},
  \bibinfo{author}{\bibfnamefont{A.}~\bibnamefont{John}}, \bibnamefont{and}
  \bibinfo{author}{\bibfnamefont{D.~D.} \bibnamefont{Seligmann}}, in
  \emph{\bibinfo{booktitle}{Computational Science and Engineering, 2009.
  CSE'09. International Conference on}} (\bibinfo{organization}{IEEE},
  \bibinfo{year}{2009}), vol.~\bibinfo{volume}{4}, pp.
  \bibinfo{pages}{151--158}.

\bibitem[{\citenamefont{Leskovec et~al.}(2009)\citenamefont{Leskovec, Lang,
  Dasgupta, and Mahoney}}]{leskovec2009community}
\bibinfo{author}{\bibfnamefont{J.}~\bibnamefont{Leskovec}},
  \bibinfo{author}{\bibfnamefont{K.~J.} \bibnamefont{Lang}},
  \bibinfo{author}{\bibfnamefont{A.}~\bibnamefont{Dasgupta}}, \bibnamefont{and}
  \bibinfo{author}{\bibfnamefont{M.~W.} \bibnamefont{Mahoney}},
  \bibinfo{journal}{Internet Mathematics} \textbf{\bibinfo{volume}{6}},
  \bibinfo{pages}{29} (\bibinfo{year}{2009}).

\bibitem[{\citenamefont{G{\'o}mez et~al.}(2008)\citenamefont{G{\'o}mez,
  Kaltenbrunner, and L{\'o}pez}}]{gomez2008statistical}
\bibinfo{author}{\bibfnamefont{V.}~\bibnamefont{G{\'o}mez}},
  \bibinfo{author}{\bibfnamefont{A.}~\bibnamefont{Kaltenbrunner}},
  \bibnamefont{and}
  \bibinfo{author}{\bibfnamefont{V.}~\bibnamefont{L{\'o}pez}}, in
  \emph{\bibinfo{booktitle}{Proceedings of the 17th international conference on
  World Wide Web}} (\bibinfo{organization}{ACM}, \bibinfo{year}{2008}), pp.
  \bibinfo{pages}{645--654}.

\bibitem[{\citenamefont{Viswanath et~al.}(2009)\citenamefont{Viswanath,
  Mislove, Cha, and Gummadi}}]{viswanath2009evolution}
\bibinfo{author}{\bibfnamefont{B.}~\bibnamefont{Viswanath}},
  \bibinfo{author}{\bibfnamefont{A.}~\bibnamefont{Mislove}},
  \bibinfo{author}{\bibfnamefont{M.}~\bibnamefont{Cha}}, \bibnamefont{and}
  \bibinfo{author}{\bibfnamefont{K.~P.} \bibnamefont{Gummadi}}, in
  \emph{\bibinfo{booktitle}{Proceedings of the 2nd ACM workshop on Online
  social networks}} (\bibinfo{organization}{ACM}, \bibinfo{year}{2009}), pp.
  \bibinfo{pages}{37--42}.

\bibitem[{\citenamefont{Richardson et~al.}(2003)\citenamefont{Richardson,
  Agrawal, and Domingos}}]{richardson2003trust}
\bibinfo{author}{\bibfnamefont{M.}~\bibnamefont{Richardson}},
  \bibinfo{author}{\bibfnamefont{R.}~\bibnamefont{Agrawal}}, \bibnamefont{and}
  \bibinfo{author}{\bibfnamefont{P.}~\bibnamefont{Domingos}}, in
  \emph{\bibinfo{booktitle}{The Semantic Web-ISWC 2003}}
  (\bibinfo{publisher}{Springer}, \bibinfo{year}{2003}), pp.
  \bibinfo{pages}{351--368}.

\bibitem[{\citenamefont{Kunegis et~al.}(2009)\citenamefont{Kunegis, Lommatzsch,
  and Bauckhage}}]{kunegis2009slashdot}
\bibinfo{author}{\bibfnamefont{J.}~\bibnamefont{Kunegis}},
  \bibinfo{author}{\bibfnamefont{A.}~\bibnamefont{Lommatzsch}},
  \bibnamefont{and}
  \bibinfo{author}{\bibfnamefont{C.}~\bibnamefont{Bauckhage}}, in
  \emph{\bibinfo{booktitle}{Proceedings of the 18th international conference on
  World wide web}} (\bibinfo{organization}{ACM}, \bibinfo{year}{2009}), pp.
  \bibinfo{pages}{741--750}.

\bibitem[{\citenamefont{McAuley and Leskovec}(2012)}]{McAuley2012}
\bibinfo{author}{\bibfnamefont{J.}~\bibnamefont{McAuley}} \bibnamefont{and}
  \bibinfo{author}{\bibfnamefont{J.}~\bibnamefont{Leskovec}}, in
  \emph{\bibinfo{booktitle}{Advances in Neural Information Processing Systems}}
  (\bibinfo{year}{2012}), pp. \bibinfo{pages}{548--556}.

\bibitem[{\citenamefont{Brandes and Lerner}(2010)}]{brandes2010structural}
\bibinfo{author}{\bibfnamefont{U.}~\bibnamefont{Brandes}} \bibnamefont{and}
  \bibinfo{author}{\bibfnamefont{J.}~\bibnamefont{Lerner}},
  \bibinfo{journal}{Journal of classification} \textbf{\bibinfo{volume}{27}},
  \bibinfo{pages}{279} (\bibinfo{year}{2010}).

\bibitem[{\citenamefont{Cho et~al.}(2011)\citenamefont{Cho, Myers, and
  Leskovec}}]{cho2011friendship}
\bibinfo{author}{\bibfnamefont{E.}~\bibnamefont{Cho}},
  \bibinfo{author}{\bibfnamefont{S.~A.} \bibnamefont{Myers}}, \bibnamefont{and}
  \bibinfo{author}{\bibfnamefont{J.}~\bibnamefont{Leskovec}}, in
  \emph{\bibinfo{booktitle}{Proceedings of the 17th ACM SIGKDD international
  conference on Knowledge discovery and data mining}}
  (\bibinfo{organization}{ACM}, \bibinfo{year}{2011}), pp.
  \bibinfo{pages}{1082--1090}.

\bibitem[{\citenamefont{Brozovsky and
  Petricek}(2007)}]{brozovsky2007recommender}
\bibinfo{author}{\bibfnamefont{L.}~\bibnamefont{Brozovsky}} \bibnamefont{and}
  \bibinfo{author}{\bibfnamefont{V.}~\bibnamefont{Petricek}},
  \bibinfo{journal}{arXiv preprint cs/0703042}  (\bibinfo{year}{2007}).

\bibitem[{\citenamefont{Kunegis et~al.}(2012)\citenamefont{Kunegis, Gr{\"o}ner,
  and Gottron}}]{kunegis2012online}
\bibinfo{author}{\bibfnamefont{J.}~\bibnamefont{Kunegis}},
  \bibinfo{author}{\bibfnamefont{G.}~\bibnamefont{Gr{\"o}ner}},
  \bibnamefont{and} \bibinfo{author}{\bibfnamefont{T.}~\bibnamefont{Gottron}},
  in \emph{\bibinfo{booktitle}{Proceedings of the 4th ACM RecSys workshop on
  Recommender systems and the social web}} (\bibinfo{organization}{ACM},
  \bibinfo{year}{2012}), pp. \bibinfo{pages}{37--44}.

\bibitem[{\citenamefont{Leskovec
  et~al.}(2007{\natexlab{b}})\citenamefont{Leskovec, Adamic, and
  Huberman}}]{leskovec2007dynamics}
\bibinfo{author}{\bibfnamefont{J.}~\bibnamefont{Leskovec}},
  \bibinfo{author}{\bibfnamefont{L.~A.} \bibnamefont{Adamic}},
  \bibnamefont{and} \bibinfo{author}{\bibfnamefont{B.~A.}
  \bibnamefont{Huberman}}, \bibinfo{journal}{ACM Transactions on the Web
  (TWEB)} \textbf{\bibinfo{volume}{1}}, \bibinfo{pages}{5}
  (\bibinfo{year}{2007}{\natexlab{b}}).

\bibitem[{\citenamefont{Albert et~al.}(1999)\citenamefont{Albert, Jeong, and
  Barab{\'a}si}}]{albert1999internet}
\bibinfo{author}{\bibfnamefont{R.}~\bibnamefont{Albert}},
  \bibinfo{author}{\bibfnamefont{H.}~\bibnamefont{Jeong}}, \bibnamefont{and}
  \bibinfo{author}{\bibfnamefont{A.-L.} \bibnamefont{Barab{\'a}si}},
  \bibinfo{journal}{Nature} \textbf{\bibinfo{volume}{401}},
  \bibinfo{pages}{130} (\bibinfo{year}{1999}).

\bibitem[{\citenamefont{Palla et~al.}(2008)\citenamefont{Palla, Farkas,
  Pollner, Der{\'e}nyi, and Vicsek}}]{palla2008fundamental}
\bibinfo{author}{\bibfnamefont{G.}~\bibnamefont{Palla}},
  \bibinfo{author}{\bibfnamefont{I.~J.} \bibnamefont{Farkas}},
  \bibinfo{author}{\bibfnamefont{P.}~\bibnamefont{Pollner}},
  \bibinfo{author}{\bibfnamefont{I.}~\bibnamefont{Der{\'e}nyi}},
  \bibnamefont{and} \bibinfo{author}{\bibfnamefont{T.}~\bibnamefont{Vicsek}},
  \bibinfo{journal}{New Journal of Physics} \textbf{\bibinfo{volume}{10}},
  \bibinfo{pages}{123026} (\bibinfo{year}{2008}).

\bibitem[{\citenamefont{Bollacker et~al.}(1998)\citenamefont{Bollacker,
  Lawrence, and Giles}}]{bollacker1998citeseer}
\bibinfo{author}{\bibfnamefont{K.~D.} \bibnamefont{Bollacker}},
  \bibinfo{author}{\bibfnamefont{S.}~\bibnamefont{Lawrence}}, \bibnamefont{and}
  \bibinfo{author}{\bibfnamefont{C.~L.} \bibnamefont{Giles}}, in
  \emph{\bibinfo{booktitle}{Proceedings of the second international conference
  on Autonomous agents}} (\bibinfo{organization}{ACM}, \bibinfo{year}{1998}),
  pp. \bibinfo{pages}{116--123}.

\bibitem[{\citenamefont{Niu et~al.}(2011)\citenamefont{Niu, Sun, Wang, Rong,
  Qi, and Yu}}]{niu2011zhishi}
\bibinfo{author}{\bibfnamefont{X.}~\bibnamefont{Niu}},
  \bibinfo{author}{\bibfnamefont{X.}~\bibnamefont{Sun}},
  \bibinfo{author}{\bibfnamefont{H.}~\bibnamefont{Wang}},
  \bibinfo{author}{\bibfnamefont{S.}~\bibnamefont{Rong}},
  \bibinfo{author}{\bibfnamefont{G.}~\bibnamefont{Qi}}, \bibnamefont{and}
  \bibinfo{author}{\bibfnamefont{Y.}~\bibnamefont{Yu}}, in
  \emph{\bibinfo{booktitle}{The Semantic Web--ISWC 2011}}
  (\bibinfo{publisher}{Springer}, \bibinfo{year}{2011}), pp.
  \bibinfo{pages}{205--220}.

\bibitem[{\citenamefont{Yang and Leskovec}(2012)}]{leskovec2012}
\bibinfo{author}{\bibfnamefont{J.}~\bibnamefont{Yang}} \bibnamefont{and}
  \bibinfo{author}{\bibfnamefont{J.}~\bibnamefont{Leskovec}}, in
  \emph{\bibinfo{booktitle}{Proceedings of the ACM SIGKDD Workshop on Mining
  Data Semantics}} (\bibinfo{organization}{ACM}, \bibinfo{year}{2012}),
  p.~\bibinfo{pages}{3}.

\bibitem[{\citenamefont{Hall et~al.}(2001)\citenamefont{Hall, Jaffe, and
  Trajtenberg}}]{hall2001nber}
\bibinfo{author}{\bibfnamefont{B.~H.} \bibnamefont{Hall}},
  \bibinfo{author}{\bibfnamefont{A.~B.} \bibnamefont{Jaffe}}, \bibnamefont{and}
  \bibinfo{author}{\bibfnamefont{M.}~\bibnamefont{Trajtenberg}},
  \bibinfo{type}{Tech. Rep.}, \bibinfo{institution}{National Bureau of Economic
  Research} (\bibinfo{year}{2001}).

\bibitem[{\citenamefont{Auer et~al.}(2007)\citenamefont{Auer, Bizer, Kobilarov,
  Lehmann, Cyganiak, and Ives}}]{auer2007dbpedia}
\bibinfo{author}{\bibfnamefont{S.}~\bibnamefont{Auer}},
  \bibinfo{author}{\bibfnamefont{C.}~\bibnamefont{Bizer}},
  \bibinfo{author}{\bibfnamefont{G.}~\bibnamefont{Kobilarov}},
  \bibinfo{author}{\bibfnamefont{J.}~\bibnamefont{Lehmann}},
  \bibinfo{author}{\bibfnamefont{R.}~\bibnamefont{Cyganiak}}, \bibnamefont{and}
  \bibinfo{author}{\bibfnamefont{Z.}~\bibnamefont{Ives}},
  \emph{\bibinfo{title}{Dbpedia: A nucleus for a web of open data}}
  (\bibinfo{publisher}{Springer}, \bibinfo{year}{2007}).

\bibitem[{\citenamefont{Mislove et~al.}(2007)\citenamefont{Mislove, Marcon,
  Gummadi, Druschel, and Bhattacharjee}}]{mislove2007measurement}
\bibinfo{author}{\bibfnamefont{A.}~\bibnamefont{Mislove}},
  \bibinfo{author}{\bibfnamefont{M.}~\bibnamefont{Marcon}},
  \bibinfo{author}{\bibfnamefont{K.~P.} \bibnamefont{Gummadi}},
  \bibinfo{author}{\bibfnamefont{P.}~\bibnamefont{Druschel}}, \bibnamefont{and}
  \bibinfo{author}{\bibfnamefont{B.}~\bibnamefont{Bhattacharjee}}, in
  \emph{\bibinfo{booktitle}{Proceedings of the 7th ACM SIGCOMM conference on
  Internet measurement}} (\bibinfo{organization}{ACM}, \bibinfo{year}{2007}),
  pp. \bibinfo{pages}{29--42}.

\end{thebibliography}

%\appendix

\begin{table*}[!htb]
\begin{center}
 \resizebox{\textwidth}{!}{%
\begin{tabular}{l|r|r|r|r|r|r|r|r|r|r} 
Network  & Ref.  & URL  & $N$  & $E$  & $\left| N^{1}_{\max} \right|$  & $\left| N^{2}_{\max} \right|$  & $T^{1}$  & $T^{2}$  & $\tilde{T}^{1}$  & $\tilde{T}^{2}$  \\ \hline 
*Social 3& \cite{milo2004superfamilies} & \href{http://wws.weizmann.ac.il/mcb/UriAlon/index.php?q=download/collection-complex-networks}{url} & $32$  & $80$  & $14$  & $22$  & $0$  & $1$  & $0$  & $0$  \\ \hline 
*Karate club& \cite{zachary1977information} & \href{http://www-personal.umich.edu/~mejn/netdata/}{url} & $34$  & $78$  & $18$  & $23$  & $0$  & $0$  & $0$  & $0$  \\ \hline 
*Protein 2& \cite{milo2004superfamilies} & \href{http://wws.weizmann.ac.il/mcb/UriAlon/index.php?q=download/collection-complex-networks}{url} & $53$  & $123$  & $8$  & $19$  & $0$  & $1$  & $0$  & $0$  \\ \hline 
*Dolphins& \cite{lusseau2003bottlenose} & \href{http://www-personal.umich.edu/~mejn/netdata/}{url} & $62$  & $159$  & $12$  & $26$  & $0$  & $3$  & $0$  & $0$  \\ \hline 
*Social 1& \cite{milo2004superfamilies} & \href{http://wws.weizmann.ac.il/mcb/UriAlon/index.php?q=download/collection-complex-networks}{url} & $67$  & $142$  & $11$  & $18$  & $0$  & $1$  & $0$  & $0$  \\ \hline 
*Les Miserables& \cite{knuth1993stanford} & \href{http://www-personal.umich.edu/~mejn/netdata/}{url} & $77$  & $254$  & $36$  & $55$  & $2$  & $9$  & $0$  & $0$  \\ \hline 
*Protein 1& \cite{milo2004superfamilies} & \href{http://wws.weizmann.ac.il/mcb/UriAlon/index.php?q=download/collection-complex-networks}{url} & $95$  & $213$  & $7$  & $12$  & $0$  & $1$  & $0$  & $1$  \\ \hline 
*E. Coli, transcription& \cite{mangan2003structure} & \href{http://wws.weizmann.ac.il/mcb/UriAlon/index.php?q=download/collection-complex-networks}{url} & $97$  & $212$  & $11$  & $18$  & $0$  & $1$  & $0$  & $1$  \\ \hline 
*Political books& \cite{adamic2005political} & \href{http://www-personal.umich.edu/~mejn/netdata/}{url} & $105$  & $441$  & $25$  & $51$  & $7$  & $62$  & $1$  & $1$  \\ \hline 
*David Copperfield& \cite{newman2006finding} & \href{http://www-personal.umich.edu/~mejn/netdata/}{url} & $112$  & $425$  & $50$  & $92$  & $13$  & $227$  & $1$  & $1$  \\ \hline 
*College football& \cite{girvan2002community} & \href{http://www-personal.umich.edu/~mejn/netdata/}{url} & $115$  & $613$  & $13$  & $39$  & $5$  & $39$  & $1$  & $1$  \\ \hline 
*S 208& \cite{milo2004superfamilies} & \href{http://wws.weizmann.ac.il/mcb/UriAlon/index.php?q=download/collection-complex-networks}{url} & $122$  & $189$  & $10$  & $10$  & $0$  & $0$  & $0$  & $0$  \\ \hline 
*High school, 2011& \cite{fournet2014contact} & \href{http://www.sociopatterns.org/datasets/high-school-dynamic-contact-networks/}{url} & $126$  & $1,709$  & $56$  & $124$  & $122$  & $1,824$  & $0$  & $0$  \\ \hline 
*Bay Dry& \cite{ulanowicz1998network, konect} & \href{http://konect.uni-koblenz.de/networks/foodweb-baydry}{url} & $128$  & $2,106$  & $111$  & $128$  & $244$  & $2,801$  & $0$  & $0$  \\ \hline 
*Bay Wet& \cite{konect} & \href{http://konect.uni-koblenz.de/networks/foodweb-baywet}{url} & $128$  & $2,075$  & $111$  & $128$  & $248$  & $2,858$  & $0$  & $0$  \\ \hline 
*Radoslaw Email& \cite{radoslaw, konect} & \href{http://konect.uni-koblenz.de/networks/radoslaw_email}{url} & $167$  & $3,250$  & $139$  & $143$  & $791$  & $2,371$  & $0$  & $0$  \\ \hline 
*High school, 2012& \cite{fournet2014contact} & \href{http://www.sociopatterns.org/datasets/high-school-dynamic-contact-networks/}{url} & $180$  & $2,220$  & $56$  & $158$  & $129$  & $4,681$  & $0$  & $0$  \\ \hline 
*Little Rock Lake& \cite{martinez1991artifacts, konect} & \href{http://konect.uni-koblenz.de/networks/maayan-foodweb}{url} & $183$  & $2,434$  & $105$  & $180$  & $413$  & $4,526$  & $0$  & $0$  \\ \hline 
*Jazz& \cite{gleiser2003community} & \href{http://deim.urv.cat/~alexandre.arenas/data/welcome.htm}{url} & $198$  & $2,742$  & $101$  & $182$  & $313$  & $7,915$  & $0$  & $0$  \\ \hline 
*S 420& \cite{milo2004superfamilies} & \href{http://wws.weizmann.ac.il/mcb/UriAlon/index.php?q=download/collection-complex-networks}{url} & $252$  & $399$  & $15$  & $14$  & $0$  & $0$  & $0$  & $0$  \\ \hline 
*C. Elegans, neural& \cite{watts1998collective} & \href{http://www-personal.umich.edu/~mejn/netdata/}{url} & $297$  & $2,148$  & $134$  & $239$  & $374$  & $15,720$  & $3$  & $3$  \\ \hline 
*Network Science& \cite{newman2006finding} & \href{http://www-personal.umich.edu/~mejn/netdata/}{url} & $379$  & $914$  & $35$  & $42$  & $4$  & $8$  & $1$  & $2$  \\ \hline 
*Dublin& \cite{isella2011s, konect} & \href{http://konect.uni-koblenz.de/networks/sociopatterns-infectious}{url} & $410$  & $2,765$  & $50$  & $164$  & $124$  & $2,342$  & $3$  & $3$  \\ \hline 
*US Air Trasportation& \cite{colizza2007reaction} & \href{https://sites.google.com/site/cxnets/usairtransportationnetwork}{url} & $500$  & $2,980$  & $146$  & $301$  & $3,119$  & $57,958$  & $3$  & $3$  \\ \hline 
*S 838& \cite{milo2004superfamilies} & \href{http://wws.weizmann.ac.il/mcb/UriAlon/index.php?q=download/collection-complex-networks}{url} & $512$  & $819$  & $23$  & $23$  & $1$  & $2$  & $1$  & $1$  \\ \hline 
*Yeast, transcription& \cite{milo2002network} & \href{http://wws.weizmann.ac.il/mcb/UriAlon/index.php?q=download/collection-complex-networks}{url} & $662$  & $1,062$  & $72$  & $80$  & $23$  & $69$  & $0$  & $1$  \\ \hline 
*URV email& \cite{guimera2003self} & \href{http://deim.urv.cat/~alexandre.arenas/data/welcome.htm}{url} & $1,133$  & $5,451$  & $72$  & $279$  & $289$  & $34,546$  & $10$  & $11$  \\ \hline 
Political blogs& \cite{adamic2005political} & \href{http://www-personal.umich.edu/~mejn/netdata/}{url} & $1,222$  & $16,714$  & $351$  & $867$  & $24,944$  & $-$  & $4$  & $4$  \\ \hline 
*Air traffic& \cite{konect} & \href{http://konect.uni-koblenz.de/networks/maayan-faa}{url} & $1,226$  & $2,408$  & $35$  & $62$  & $28$  & $139$  & $5$  & $8$  \\ \hline 
*Yeast, protein& \cite{jeong2001lethality} & \href{http://www3.nd.edu/~networks/resources.htm}{url} & $1,458$  & $1,948$  & $57$  & $57$  & $16$  & $34$  & $2$  & $3$  \\ \hline 
Petster, hamster& \cite{konect} & \href{http://konect.uni-koblenz.de/networks/petster-friendships-hamster}{url} & $1,788$  & $12,476$  & $272$  & $904$  & $24,879$  & $-$  & $25$  & $23$  \\ \hline 
UC Irvine& \cite{opsahl2009clustering, konect} & \href{http://konect.uni-koblenz.de/networks/opsahl-ucsocial}{url} & $1,893$  & $13,835$  & $256$  & $1,076$  & $11,963$  & $-$  & $9$  & $10$  \\ \hline 
Yeast, protein& \cite{bu2003topological} & \href{http://vlado.fmf.uni-lj.si/pub/networks/data/bio/Yeast/Yeast.htm}{url} & $2,224$  & $6,609$  & $65$  & $208$  & $-$  & $-$  & $9$  & $13$  \\ \hline 
Japanese& \cite{milo2004superfamilies} & \href{http://wws.weizmann.ac.il/mcb/UriAlon/index.php?q=download/collection-complex-networks}{url} & $2,698$  & $7,995$  & $726$  & $1,459$  & $181,172$  & $-$  & $8$  & $10$  \\ \hline 
Open flights& \cite{opsahl2010node, konect} & \href{http://konect.uni-koblenz.de/networks/opsahl-openflights}{url} & $2,905$  & $15,645$  & $242$  & $1,101$  & $-$  & $-$  & $18$  & $18$  \\ \hline 
*GR-QC, 1993-2003& \cite{leskovec2007graph} & \href{http://snap.stanford.edu/data/ca-GrQc.html}{url} & $4,158$  & $13,422$  & $81$  & $187$  & $356$  & $10,098$  & $31$  & $40$  \\ \hline 
Tennis& \cite{radicchi2011best} & \href{-}{url} & $4,338$  & $81,865$  & $452$  & $2,015$  & $-$  & $-$  & $21$  & $23$  \\ \hline 
\end{tabular}
}
\end{center}
\caption{From left to right, we report: the name of the network, the
  reference of the paper(s) where the network was first analyzed, the
  URL where the network was retrieved,
  the number of nodes and edges of the network, the number of
  nodes in the largest neighborhoods of orders $r=1$ and
  $r=2$,
  and the computational time needed to obtain the spectral
  density of the graph Laplacian via the NMP approximation
  with $r_{\max}=1$ and $K=N$,   $r_{\max}=2$ and $K=N$, $r_{\max}=1$ and
  $K=10$, and $r_{\max}=2$ and $K=10$.
  The analysis was performed on the largest
  connected component of each network.
  Computational time is measured in seconds, and the reported value is
  rounded to the nearest integer. A computational time equal to zero seconds means that less than
  $0.5$ seconds were required to estimate the spectral density of the
  graph Laplacian. No computational time is reported for networks that
  could not be analyzed due to their high computational
  demand. The asterisk before the network name indicate that we were
  able to fully compute their spectrum, either exactly or using the
  various NMP approximations. These are the only networks
  included in the analysis of Figs.~\ref{fig:6} and ~\ref{fig:7}.}
\label{tab:1}
\end{table*}

\begin{table*}[!htb]
\begin{center}
 \resizebox{\textwidth}{!}{%
\begin{tabular}{l|r|r|r|r|r|r|r|r|r|r} 
Network  & Ref.  & URL  & $N$  & $E$  & $\left| N^{1}_{\max} \right|$  & $\left| N^{2}_{\max} \right|$  & $T^{1}$  & $T^{2}$  & $\tilde{T}^{1}$  & $\tilde{T}^{2}$  \\ \hline 
US Power grid& \cite{watts1998collective} & \href{http://www-personal.umich.edu/~mejn/netdata/}{url} & $4,941$  & $6,594$  & $20$  & $28$  & $-$  & $-$  & $13$  & $19$  \\ \hline 
HT09& \cite{isella2011s} & \href{http://www.sociopatterns.org/datasets/hypertext-2009-dynamic-contact-network/}{url} & $5,352$  & $18,481$  & $1,288$  & $1,464$  & $-$  & $-$  & $4$  & $9$  \\ \hline 
Hep-Th, 1995-1999& \cite{newman2001structure} & \href{http://www-personal.umich.edu/~mejn/netdata/}{url} & $5,835$  & $13,815$  & $50$  & $136$  & $-$  & $-$  & $36$  & $53$  \\ \hline 
Reactome& \cite{joshi2005reactome, konect} & \href{http://konect.uni-koblenz.de/networks/reactome}{url} & $5,973$  & $145,778$  & $855$  & $2,492$  & $-$  & $-$  & $135$  & $129$  \\ \hline 
Jung& \cite{vsubelj2012software, konect} & \href{http://konect.uni-koblenz.de/networks/subelj_jung-j}{url} & $6,120$  & $50,290$  & $5,656$  & $6,050$  & $-$  & $-$  & $137$  & $250$  \\ \hline 
Gnutella, Aug. 8, 2002& \cite{ripeanu2002mapping, leskovec2007graph} & \href{http://snap.stanford.edu/data/p2p-Gnutella08.html}{url} & $6,299$  & $20,776$  & $98$  & $440$  & $-$  & $-$  & $14$  & $26$  \\ \hline 
JDK& \cite{konect} & \href{http://konect.uni-koblenz.de/networks/subelj_jdk}{url} & $6,434$  & $53,658$  & $5,923$  & $6,356$  & $-$  & $-$  & $151$  & $288$  \\ \hline 
AS Oregon& \cite{leskovec2005graphs} & \href{http://snap.stanford.edu/data/as.html}{url} & $6,474$  & $12,572$  & $1,459$  & $2,685$  & $-$  & $-$  & $10$  & $18$  \\ \hline 
English& \cite{milo2004superfamilies} & \href{http://wws.weizmann.ac.il/mcb/UriAlon/index.php?q=download/collection-complex-networks}{url} & $7,377$  & $44,205$  & $2,569$  & $5,585$  & $-$  & $-$  & $47$  & $69$  \\ \hline 
Gnutella, Aug. 9, 2002& \cite{ripeanu2002mapping, leskovec2007graph} & \href{http://snap.stanford.edu/data/p2p-Gnutella09.html}{url} & $8,104$  & $26,008$  & $103$  & $421$  & $-$  & $-$  & $16$  & $31$  \\ \hline 
French& \cite{milo2004superfamilies} & \href{http://wws.weizmann.ac.il/mcb/UriAlon/index.php?q=download/collection-complex-networks}{url} & $8,308$  & $23,832$  & $1,892$  & $4,405$  & $-$  & $-$  & $30$  & $50$  \\ \hline 
Hep-Th, 1993-2003& \cite{leskovec2007graph} & \href{http://snap.stanford.edu/data/ca-HepTh.html}{url} & $8,638$  & $24,806$  & $65$  & $244$  & $-$  & $-$  & $64$  & $82$  \\ \hline 
Gnutella, Aug. 6, 2002& \cite{ripeanu2002mapping, leskovec2007graph} & \href{http://snap.stanford.edu/data/p2p-Gnutella06.html}{url} & $8,717$  & $31,525$  & $115$  & $266$  & $-$  & $-$  & $20$  & $40$  \\ \hline 
Gnutella, Aug. 5, 2002& \cite{ripeanu2002mapping, leskovec2007graph} & \href{http://snap.stanford.edu/data/p2p-Gnutella05.html}{url} & $8,842$  & $31,837$  & $89$  & $321$  & $-$  & $-$  & $22$  & $42$  \\ \hline 
PGP& \cite{boguna2004models} & \href{http://deim.urv.cat/~alexandre.arenas/data/welcome.htm}{url} & $10,680$  & $24,316$  & $206$  & $481$  & $-$  & $-$  & $74$  & $63$  \\ \hline 
Gnutella, August 4 2002& \cite{ripeanu2002mapping, leskovec2007graph} & \href{http://snap.stanford.edu/data/p2p-Gnutella04.html}{url} & $10,876$  & $39,994$  & $103$  & $250$  & $-$  & $-$  & $30$  & $47$  \\ \hline 
Hep-Ph, 1993-2003& \cite{leskovec2007graph} & \href{http://snap.stanford.edu/data/ca-HepPh.html}{url} & $11,204$  & $117,619$  & $491$  & $1,960$  & $-$  & $-$  & $247$  & $242$  \\ \hline 
Spanish& \cite{milo2004superfamilies} & \href{http://wws.weizmann.ac.il/mcb/UriAlon/index.php?q=download/collection-complex-networks}{url} & $11,558$  & $43,050$  & $2,986$  & $7,814$  & $-$  & $-$  & $59$  & $75$  \\ \hline 
DBLP, citations& \cite{ley2002dblp, konect} & \href{http://konect.uni-koblenz.de/networks/dblp-cite}{url} & $12,495$  & $49,563$  & $710$  & $2,876$  & $-$  & $-$  & $67$  & $77$  \\ \hline 
Spanish& \cite{konect} & \href{http://konect.uni-koblenz.de/networks/lasagne-spanishbook}{url} & $12,643$  & $55,019$  & $5,170$  & $11,524$  & $-$  & $-$  & $75$  & $117$  \\ \hline 
*Cond-Mat, 1995-1999& \cite{newman2001structure} & \href{http://www-personal.umich.edu/~mejn/netdata/}{url} & $13,861$  & $44,619$  & $107$  & $358$  & $1,478$  & $98,257$  & $167$  & $188$  \\ \hline 
Astrophysics& \cite{newman2001structure} & \href{http://www-personal.umich.edu/~mejn/netdata/}{url} & $14,845$  & $119,652$  & $361$  & $2,050$  & $-$  & $-$  & $187$  & $212$  \\ \hline 
Google& \cite{palla2007directed} & \href{http://cfinder.org}{url} & $15,763$  & $148,585$  & $11,401$  & $13,208$  & $-$  & $-$  & $455$  & $740$  \\ \hline 
AstroPhys, 1993-2003& \cite{leskovec2007graph} & \href{http://snap.stanford.edu/data/ca-AstroPh.html}{url} & $17,903$  & $196,972$  & $504$  & $3,661$  & $-$  & $-$  & $219$  & $240$  \\ \hline 
Cond-Mat, 1993-2003& \cite{leskovec2007graph} & \href{http://snap.stanford.edu/data/ca-CondMat.html}{url} & $21,363$  & $91,286$  & $280$  & $1,335$  & $-$  & $-$  & $433$  & $582$  \\ \hline 
Gnutella, Aug. 25, 2002& \cite{ripeanu2002mapping, leskovec2007graph} & \href{http://snap.stanford.edu/data/p2p-Gnutella25.html}{url} & $22,663$  & $54,693$  & $67$  & $85$  & $-$  & $-$  & $42$  & $59$  \\ \hline 
Internet& - & \href{http://www-personal.umich.edu/~mejn/netdata/}{url} & $22,963$  & $48,436$  & $2,390$  & $6,954$  & $-$  & $-$  & $62$  & $82$  \\ \hline 
Thesaurus& \cite{kiss1973associative, konect} & \href{http://konect.uni-koblenz.de/networks/eat}{url} & $23,132$  & $297,094$  & $1,062$  & $9,528$  & $-$  & $-$  & $165$  & $217$  \\ \hline 
Cora& \cite{vsubelj2013model, konect} & \href{http://konect.uni-koblenz.de/networks/subelj_cora}{url} & $23,166$  & $89,157$  & $377$  & $818$  & $-$  & $-$  & $-$  & $-$  \\ \hline 
Linux, mailing list& \cite{konect} & \href{http://konect.uni-koblenz.de/networks/lkml-reply}{url} & $24,567$  & $158,164$  & $2,989$  & $10,805$  & $-$  & $-$  & $520$  & $622$  \\ \hline 
AS Caida& \cite{leskovec2005graphs} & \href{http://snap.stanford.edu/data/as-caida.html}{url} & $26,475$  & $53,381$  & $2,629$  & $7,940$  & $-$  & $-$  & $68$  & $97$  \\ \hline 
Gnutella, Aug. 24, 2002& \cite{ripeanu2002mapping, leskovec2007graph} & \href{http://snap.stanford.edu/data/p2p-Gnutella24.html}{url} & $26,498$  & $65,359$  & $355$  & $860$  & $-$  & $-$  & $-$  & $-$  \\ \hline 
Hep-Th, citations& \cite{leskovec2007graph, konect} & \href{http://konect.uni-koblenz.de/networks/cit-HepTh}{url} & $27,400$  & $352,021$  & $2,469$  & $9,467$  & $-$  & $-$  & $-$  & $-$  \\ \hline 
Cond-Mat, 1995-2003& \cite{newman2001structure} & \href{http://www-personal.umich.edu/~mejn/netdata/}{url} & $27,519$  & $116,181$  & $202$  & $1,108$  & $-$  & $-$  & $648$  & $672$  \\ \hline 
Digg& \cite{de2009social, konect} & \href{http://konect.uni-koblenz.de/networks/munmun_digg_reply}{url} & $29,652$  & $84,781$  & $283$  & $1,657$  & $-$  & $-$  & $189$  & $286$  \\ \hline 
Linux, soft.& \cite{konect} & \href{http://konect.uni-koblenz.de/networks/linux}{url} & $30,817$  & $213,208$  & $9,339$  & $18,740$  & $-$  & $-$  & $1,195$  & $1,759$  \\ \hline 
Enron& \cite{leskovec2009community} & \href{http://snap.stanford.edu/data/email-Enron.html}{url} & $33,696$  & $180,811$  & $1,383$  & $8,264$  & $-$  & $-$  & $590$  & $550$  \\ \hline 
\end{tabular}
}
\end{center}
\caption{Continuation of Table~\ref{tab:1}.}
\label{tab:2}
\end{table*}

\begin{table*}[!htb]
\begin{center}
 \resizebox{\textwidth}{!}{%
\begin{tabular}{l|r|r|r|r|r|r|r|r|r|r} 
Network  & Ref.  & URL  & $N$  & $E$  & $\left| N^{1}_{\max} \right|$  & $\left| N^{2}_{\max} \right|$  & $T^{1}$  & $T^{2}$  & $\tilde{T}^{1}$  & $\tilde{T}^{2}$  \\ \hline 
Hep-Ph, citations& \cite{leskovec2007graph, konect} & \href{http://konect.uni-koblenz.de/networks/cit-HepPh}{url} & $34,401$  & $420,784$  & $846$  & $4,440$  & $-$  & $-$  & $-$  & $-$  \\ \hline 
Cond-Mat, 1995-2005& \cite{newman2001structure} & \href{http://www-personal.umich.edu/~mejn/netdata/}{url} & $36,458$  & $171,735$  & $278$  & $1,855$  & $-$  & $-$  & $899$  & $948$  \\ \hline 
Gnutella, Aug. 30, 2002& \cite{ripeanu2002mapping, leskovec2007graph} & \href{http://snap.stanford.edu/data/p2p-Gnutella30.html}{url} & $36,646$  & $88,303$  & $56$  & $86$  & $-$  & $-$  & $-$  & $-$  \\ \hline 
Slashdot& \cite{gomez2008statistical, konect} & \href{http://konect.uni-koblenz.de/networks/slashdot-threads}{url} & $51,083$  & $116,573$  & $2,916$  & $7,732$  & $-$  & $-$  & $-$  & $-$  \\ \hline 
Gnutella, Aug. 31, 2002& \cite{ripeanu2002mapping, leskovec2007graph} & \href{http://snap.stanford.edu/data/p2p-Gnutella31.html}{url} & $62,561$  & $147,878$  & $96$  & $111$  & $-$  & $-$  & $-$  & $-$  \\ \hline 
Facebook& \cite{viswanath2009evolution} & \href{http://socialnetworks.mpi-sws.org/data-wosn2009.html}{url} & $63,392$  & $816,886$  & $1,099$  & $14,291$  & $-$  & $-$  & $-$  & $-$  \\ \hline 
Epinions& \cite{richardson2003trust, konect} & \href{http://konect.uni-koblenz.de/networks/soc-Epinions1}{url} & $75,877$  & $405,739$  & $3,045$  & $16,661$  & $-$  & $-$  & $-$  & $-$  \\ \hline 
Slashdot zoo& \cite{kunegis2009slashdot, konect} & \href{http://konect.uni-koblenz.de/networks/slashdot-zoo}{url} & $79,116$  & $467,731$  & $2,534$  & $16,810$  & $-$  & $-$  & $-$  & $-$  \\ \hline 
Flickr& \cite{McAuley2012, konect} & \href{http://konect.uni-koblenz.de/networks/flickrEdges}{url} & $105,722$  & $2,316,668$  & $5,425$  & $15,360$  & $-$  & $-$  & $-$  & $-$  \\ \hline 
Wikipedia, edits& \cite{brandes2010structural, konect} & \href{http://konect.uni-koblenz.de/networks/wikiconflict}{url} & $113,123$  & $2,025,910$  & $20,153$  & $72,317$  & $-$  & $-$  & $-$  & $-$  \\ \hline 
Petster, cats& \cite{konect} & \href{http://konect.uni-koblenz.de/networks/petster-friendships-cat}{url} & $148,826$  & $5,447,464$  & $80,634$  & $136,538$  & $-$  & $-$  & $-$  & $-$  \\ \hline 
Gowalla& \cite{cho2011friendship, konect} & \href{http://konect.uni-koblenz.de/networks/loc-gowalla_edges}{url} & $196,591$  & $950,327$  & $14,730$  & $55,087$  & $-$  & $-$  & $-$  & $-$  \\ \hline 
Libimseti& \cite{brozovsky2007recommender, kunegis2012online, konect} & \href{http://konect.uni-koblenz.de/networks/libimseti}{url} & $220,970$  & $17,233,144$  & $33,390$  & $181,596$  & $-$  & $-$  & $-$  & $-$  \\ \hline 
EU email& \cite{leskovec2007graph, konect} & \href{http://konect.uni-koblenz.de/networks/email-EuAll}{url} & $224,832$  & $339,925$  & $7,636$  & $20,391$  & $-$  & $-$  & $-$  & $-$  \\ \hline 
Web Stanford& \cite{leskovec2009community} & \href{http://snap.stanford.edu/data/web-Stanford.html}{url} & $255,265$  & $1,941,926$  & $38,625$  & $54,427$  & $-$  & $-$  & $-$  & $-$  \\ \hline 
Amazon, Mar. 2, 2003& \cite{leskovec2007dynamics} & \href{http://snap.stanford.edu/data/amazon0302.html}{url} & $262,111$  & $899,792$  & $420$  & $944$  & $-$  & $-$  & $-$  & $-$  \\ \hline 
DBLP, collaborations& \cite{ley2002dblp, konect} & \href{http://konect.uni-koblenz.de/networks/dblp_coauthor}{url} & $317,080$  & $1,049,866$  & $344$  & $1,431$  & $-$  & $-$  & $-$  & $-$  \\ \hline 
Web Notre Dame& \cite{albert1999internet} & \href{http://www3.nd.edu/~networks/resources.htm}{url} & $325,729$  & $1,090,108$  & $10,722$  & $17,682$  & $-$  & $-$  & $-$  & $-$  \\ \hline 
MathSciNet& \cite{palla2008fundamental} & \href{http://cfinder.org}{url} & $332,689$  & $820,644$  & $496$  & $2,454$  & $-$  & $-$  & $-$  & $-$  \\ \hline 
CiteSeer& \cite{bollacker1998citeseer, konect} & \href{http://konect.uni-koblenz.de/networks/citeseer}{url} & $365,154$  & $1,721,981$  & $1,739$  & $5,392$  & $-$  & $-$  & $-$  & $-$  \\ \hline 
Zhishi& \cite{niu2011zhishi, konect} & \href{http://konect.uni-koblenz.de/networks/zhishi-baidu-relatedpages}{url} & $372,840$  & $2,318,025$  & $127,067$  & $128,431$  & $-$  & $-$  & $-$  & $-$  \\ \hline 
Actor coll. net.& \cite{barabasi1999emergence, konect} & \href{http://konect.uni-koblenz.de/networks/actor-collaboration}{url} & $374,511$  & $15,014,839$  & $3,956$  & $125,645$  & $-$  & $-$  & $-$  & $-$  \\ \hline 
Amazon, Mar. 12, 2003& \cite{leskovec2007dynamics} & \href{http://snap.stanford.edu/data/amazon0312.html}{url} & $400,727$  & $2,349,869$  & $2,747$  & $6,158$  & $-$  & $-$  & $-$  & $-$  \\ \hline 
Amazon, Jun. 6, 2003& \cite{leskovec2007dynamics} & \href{http://snap.stanford.edu/data/amazon0601.html}{url} & $403,364$  & $2,443,311$  & $2,752$  & $5,738$  & $-$  & $-$  & $-$  & $-$  \\ \hline 
Amazon, May 5, 2003& \cite{leskovec2007dynamics} & \href{http://snap.stanford.edu/data/amazon0505.html}{url} & $410,236$  & $2,439,437$  & $2,760$  & $6,491$  & $-$  & $-$  & $-$  & $-$  \\ \hline 
Petster, dogs& \cite{konect} & \href{http://konect.uni-koblenz.de/networks/petster-friendships-dog}{url} & $426,485$  & $8,543,321$  & $46,504$  & $313,475$  & $-$  & $-$  & $-$  & $-$  \\ \hline 
Road network PA& \cite{leskovec2009community} & \href{http://snap.stanford.edu/data/roadNet-PA.html}{url} & $1,087,562$  & $1,541,514$  & $9$  & $13$  & $-$  & $-$  & $-$  & $-$  \\ \hline 
YouTube friend. net.& \cite{leskovec2012, konect} & \href{http://konect.uni-koblenz.de/networks/com-youtube}{url} & $1,134,890$  & $2,987,624$  & $28,755$  & $137,387$  & $-$  & $-$  & $-$  & $-$  \\ \hline 
Road network TX& \cite{leskovec2009community} & \href{http://snap.stanford.edu/data/roadNet-TX.html}{url} & $1,351,137$  & $1,879,201$  & $13$  & $19$  & $-$  & $-$  & $-$  & $-$  \\ \hline 
AS Skitter& \cite{leskovec2005graphs} & \href{http://snap.stanford.edu/data/as-skitter.html}{url} & $1,694,616$  & $11,094,209$  & $35,455$  & $128,203$  & $-$  & $-$  & $-$  & $-$  \\ \hline 
Road network CA& \cite{leskovec2009community} & \href{http://snap.stanford.edu/data/roadNet-CA.html}{url} & $1,957,027$  & $2,760,388$  & $13$  & $17$  & $-$  & $-$  & $-$  & $-$  \\ \hline 
Wikipedia, pages& \cite{palla2008fundamental} & \href{http://cfinder.org}{url} & $2,070,367$  & $42,336,614$  & $230,041$  & $1,640,275$  & $-$  & $-$  & $-$  & $-$  \\ \hline 
US Patents& \cite{hall2001nber, konect} & \href{http://konect.uni-koblenz.de/networks/patentcite}{url} & $3,764,117$  & $16,511,740$  & $794$  & $4,228$  & $-$  & $-$  & $-$  & $-$  \\ \hline 
DBpedia& \cite{auer2007dbpedia, konect} & \href{http://konect.uni-koblenz.de/networks/dbpedia-all}{url} & $3,915,921$  & $12,577,253$  & $469,692$  & $897,744$  & $-$  & $-$  & $-$  & $-$  \\ \hline 
LiveJournal& \cite{mislove2007measurement, konect} & \href{http://konect.uni-koblenz.de/networks/livejournal-links}{url} & $5,189,809$  & $48,688,097$  & $15,018$  & $182,439$  & $-$  & $-$  & $-$  & $-$  \\ \hline 
\end{tabular}
}
\end{center}
\caption{Continuation of Tables~\ref{tab:1} and~\ref{tab:2}.}
\label{tab:3}
\end{table*}

%% 

%\include{table}

\begin{comment}

\caption{From left to right, we report: the name of the network, the
  reference of the paper(s) where the network was first analyzed, the
  URL where the network was retrieved,
  the number of nodes and edges of the network, the number of
  nodes and edges in the largest neighborhoods of orders $r=1$ and
  $r=2$,
  and the computational time needed to obtain the spectral
  density of the graph Laplacian via the message-passing approximation
  with $r_\max=1$ and $K=N$,   $r_\max=2$ and $K=N$, $r_\max=1$ and
  $K=10$, and $r_\max=2$ and $K=10$.
  The analysis was performed on the largest
  connected component of each network.
  Computational time is measured in seconds, and the reported value is
  rounded to the nearest integer. A computational time equal to zero seconds means that less than
  $0.5$ seconds were required to estimate the spectral density of the
  graph Laplacian. No computational time is reported for networks that
  could not be analyzed due to their high computational
  demand. The asterisk before the network name indicate that we were
  able to fully compute their spectrum, either exactly or using the
  various message-passing approximations. These are the only networks
  included in the analysis of Fig.~\ref{fig:6}.
}

\caption{Continuation of Table~\ref{tab:1}.}

\caption{Continuation of Tables~\ref{tab:1} and~\ref{tab:2}.}

\end{comment}

\end{document}